\documentclass[aps,prx,twocolumn,superscriptaddress,showpacs,floatfix,longbibliography]{revtex4-1}
\usepackage{amsmath,amssymb,graphics,epsfig,epstopdf,color,verbatim,braket,tabularx}
\usepackage{multirow}
\usepackage{dcolumn}
\usepackage{bm}
\usepackage{amssymb}
\usepackage{supertabular}
\usepackage[colorlinks,linkcolor=blue,citecolor=blue,urlcolor=blue,bookmarks=false,breaklinks=true]{hyperref}
\begin{document}

\title{Topological metals induced by Zeeman effect}

\author{Song Sun}
\affiliation{Beijing National Laboratory for Condensed Matter Physics, and Institute of Physics, Chinese Academy of Sciences, Beijing 100190, China}
\affiliation{University of Chinese Academy of Sciences, Beijing 100049, China}

\author{Zhida Song}
\affiliation{Beijing National Laboratory for Condensed Matter Physics, and Institute of Physics, Chinese Academy of Sciences, Beijing 100190, China}
\affiliation{University of Chinese Academy of Sciences, Beijing 100049, China}

\author{Hongming Weng}
\affiliation{Beijing National Laboratory for Condensed Matter Physics, and Institute of Physics, Chinese Academy of Sciences, Beijing 100190, China}
\affiliation{University of Chinese Academy of Sciences, Beijing 100049, China}
\affiliation{Collaborative Innovation Center of Quantum Matter, Beijing, China}

\author{Xi Dai}
\email{daix@ust.hk}
\affiliation{Department of Physics, Hong Kong University of Science and Technology, Clear Water Bay, Hong Kong}

\begin{abstract}

In the present paper, we propose a new way to classify centrosymmetric metals by studying the Zeeman effect caused by an external magnetic field described by the momentum dependent g-factor tensor on the Fermi surfaces. 
Nontrivial U(1) Berry's phase and curvature can be generated once the otherwise degenerate Fermi surfaces are splitted by the Zeeman effect, which will be determined by both the intrinsic band structure and the structure of g-factor tensor on the manifold of the Fermi surfaces. 
Such Zeeman effect generated Berry's phase and curvature can lead to three important experimental effects, modification of spin-zero effect, Zeeman effect induced Fermi surface Chern number and the in-plane anomalous Hall effect.
By first principle calculations, we study all these effects on two typical material, ZrTe$_5$ and TaAs$_2$ and the results are in good agreement with the existing experiments.

\end{abstract}

\maketitle

\section{Introduction}

How does a condensed matter system response to an external magnetic field is one of the key properties signaling its low energy electronic structure. For metallic systems, most of the magnetic responses, such as quantum oscillation spectrum, magneto resistance and Hall effects, are determined by the Bloch states at the Fermi surfaces(FS) only, where in nonmagnetic centrosymmetric metals the effects caused by the magnetic field can be ascribed to two types, the Zeeman effect that splits the otherwise degenerate bands and the orbital effect that leads to Landau levels.  
Without magnetic field, due to the time reversal and inversion symmetries, the Fermi surfaces always have two fold degeneracy and the Berry's connection in this case is SU(2) and traceless. Once a magnetic field is applied, the Zeeman effect described by a momentum dependent 2$\times$2 g-factor tensor will split the doubly degenerate FS into two separate ones, with each one of them being treated as a non-degenerate system carrying ordinary U(1) Berry's connection (curvature), which is crucial to determine the low energy dynamics of the Bloch electrons around the FS. Most importantly, the Zeeman effect induced U(1) Berry's curvature could be very large in metals with large spin orbital coupling and leads to several interesting physical phenomena, which have been overlooked for decades. 

In general, for such systems the Zeeman effect are described by the g-factor tensor which can be expressed as a momentum dependent vector valued 2$\times$2 matrix $\bm{\hat{g}}(\bm{k})$. As discussed in detail previously\cite{cohen_g-factor_1960,luttinger_perturbation,song_first_principle}, the re-normalization of the g-factor tensor from its vacuum value as well as its k-dependence are both caused by the high energy bands through the down folding process, or equivalently due to the effective screening process for the diamagnetic current.  After down folding to the lowest bands forming the FS, the Zeeman effect induced Berry’s connection can be fully determined by the g factor tensor $\bm{\hat{g}}(\bm{k})$ and the original SU(2) non-Abelian Berry’s connection $\bm{\mathcal{\hat{A}}}(\bm{k})$ on the FS.

Such Zeeman effect induced Berry’s connection can lead to many interesting physical phenomena. 
First of all, the induced Berry’s connection will contribute a phase to the quantization condition for Landau orbits, which can be detected directly by the quantum oscillation spectrum\cite{mikitik1999manifestation,zhang2005experimental,Novoselov2006}. 
As will be introduced below, such additional phase will strongly modify the ``spin-zero” effect\cite{shoenberg_magnetic_1984} in these materials, in which the amplitude of quantum oscillation will vanish completely at some special angles determined by the Zeeman effect.  
In traditional materials, the ``spin-zero" angles are fully determined by the splitting of the areas enclosed by Landau orbits for two different ``spin"\cite{shoenberg_magnetic_1984}. 
While, for metals with strong spin orbital coupling these ``spin-zero" angles acquired considerable contributions from Zeeman induced Berry's phases accumulated on two different Landau orbits, which will greatly change the ``spin-zero" angles. 
Secondly, the Zeeman induced Berry’s curvature will contribute to the Hall effect in addition to the ordinary Lorentz force, which has the same origin with the anomalous Hall effect\cite{RevModPhys.82.1539,FangAHEScience,AHEinFerroma} in the ferromagnetic metals. When the magnetic field is applied within the plane, the Lorentz force can be neglected and such a in-plane anomalous Hall effect is mainly contributed by the Zeeman effect induced Berry’s curvature and can be quite pronounced in some materials as shown below. 
Last, most interestingly, such Zeeman induced Berry’s curvature can be integrated over each particular closed FS and leads to topological metal phase\cite{WanPhysRevB.83.205101,Type2Weyl,WangNa3Bi,Liu2014Dirac,WengTaAsPRX,ExpTaAsPRX,Xu613,KimDiracLineNodes,Schoop2016DiracLineNode,RevModPhys.90.015001} under the magnetic field if such integrations reach nonzero integers. 
In such cases, these otherwise degenerate FS will split into two with opposite nonzero Chern numbers, which will lead to similar chiral magnetic effect\cite{fukushima2008chiral,CMEZrTr5,CMEson} and negative magneto-resistance\cite{abj_1983,PhysRevX.5.031023,son_NMR_2013} as those Weyl semimetals. 

In the present paper, we will first introduce the general theory for the Zeeman induced Berry's connection in nonmagnetic centrosymmetric metals. After that, we will take ZrTe$_5$ and TaAs$_2$ as two typical examples to introduce the ``spin-zero" effect, in-plane anomalous Hall effect and the field induced topological metals in these material systems.

\section{Theory}\label{Theory}
In nonmagnetic centrosymmetric systems, the Bloch states are always doubly degenerate at any $\bm{k}$ point, which is guaranteed by the combination of time reversal and space inversion symmetry $\mathcal{PT}$. There is a SU(2) gauge freedom stemming from the degenerate subspace at each $\bm{k}$ point, and the non-Abelian traceless Berry connection $\bm{\mathcal{\hat{A}}}_{aa'}(\bm{k})=i\langle\psi_a|\partial_{\bm{k}}\psi_{a'}\rangle$ and Berry curvature $\bm{\hat{\mathcal{F}}}(\bm{k})=\nabla_{\bm{k}}\times \bm{\mathcal{\hat{A}}}-i\bm{\mathcal{\hat{A}}}\times\bm{\mathcal{\hat{A}}}$ can be defined, which determines the low energy dynamics for the quasi-particles near the Fermi level.
Under an U(2) gauge transformation $\hat{U}(\bm{k})$, the above Berry's connection and curvature are transformed in the following way\cite{RevModPhys.Niu}: $\bm{\mathcal{\hat A}}'= {\hat U}^\dagger\bm{\mathcal{\hat A}}{\hat U}+i{\hat U}^{\dagger}\partial_{\bm{k}}{\hat U}$, $\bm{\hat{\mathcal{F}}}'=U^\dagger\bm{\hat{\mathcal{F}}}U$. 
When magnetic field is applied, the Zeeman's coupling $\hat{H}^z_{aa'}=\mu_B \bm{\hat{g}}_{aa'}(\bm{k})\cdot\bm{B} $ will break the time reversal symmetry and split the degenerate states. 
In this case, the Berry connection reduces to U(1) and can be obtained from the the previous SU(2) Berry's connection $\bm{\mathcal{\hat{A}}}(\bm{k})$ and a specific SU(2) matrix $\hat U(\bm{k})$ which diagonalize the Zeeman's coupling at each particular $\bm{k}$ point as $\bm{A}_{\pm}=\pm \frac{1}{2}\mathrm{Tr}[\hat{U}^\dagger\bm{\mathcal{\hat{A}}}\hat{U}{\hat\sigma_z}+i\hat{U}^{\dagger}\partial_{\bm{k}}\hat{U}{\hat\sigma_z}]$ with the $\pm$ sign representing the two branches of bands after splitting. 
Since the original SU(2) Berry's connection  $\bm{\mathcal{\hat{A}}}(\bm{k})$ is traceless, we always have  $\bm{A}_+=-\bm{A}_{-}$. Details about the properties of SU(2) Berry's connection under $\mathcal{PT}$ symmetry are given in Appendix \ref{sec:PT}.
It is worth emphasizing that the U(1) Berry's connection  $\bm{A}_{\pm}(\bm{k})$ is determined not only by the topological features of the degenerate band structure before the Zeeman splitting, the SU(2) Berry's connection  $\bm{\mathcal{\hat{A}}}(\bm{k})$,  but also by the topological structure hidden in the k-dependent g-factor $\bm{\hat{g}}(\bm{k}) $. Therefore, to determine the topological features of a metallic system with Zeeman splitted FS, the dependence of the g-factor on the FS is essential.

In solid state systems, not only the spin but also the orbital responses contribute to the g-factor. The spin contribution $\bm{\hat{g}}^{s}$ can be calculated directly from the corresponding Bloch functions. In our previous paper\cite{song_first_principle}, we have already developed the computational method to compute the orbital contribution $\bm{\hat{g}}^{o}$, which will be briefly sketched here.
Around the wave vector $\bm{K}$, the ``bare" k$\cdot$p Hamiltonian has the form \cite{winkler_spin--orbit_2003}
\begin{equation}
   \hat{H}_{nn'}=\delta_{nn'}\left(\epsilon_n+\frac{\hbar^2\bm{k}^2}{2m_\mathrm{e}}\right)+\bm{\hat{v}}_{nn'}\cdot\bm{k} 
\end{equation}
An unitary transformation can be obtained by the quasi-degenerate perturbation to decouple the subspace we focus on (called ``low-energy subspace") from all the other bands (called ``high-energy subspace")\cite{lowdin1951note,winkler_spin--orbit_2003}
\begin{align}\label{eq:downfoldH}
   \hat{H}&_{mm'}=\delta_{mm'}\left(\epsilon_m+\frac{\hbar^2\bm{k}^2}{2m_\mathrm{e}}\right)+\bm{\hat{v}}_{mm'}\cdot\bm{k} \nonumber\\
   &+\frac{1}{2}\sum_{l,ij}\left(\frac{1}{\epsilon_{m}-\epsilon_{l}}+\frac{1}{\epsilon_{m'}-\epsilon_{l}}\right) \hat{v}^{i}_{ml} \hat{v}^{j}_{lm'} k^i k^j
\end{align}
, where the index $m$,$m'$ are the band indexes for the low-energy subspace, and $l$ is the band index for the high-energy subspace. In the presence of magnetic field, according to Perierls substitution the momentums $k^i$ should be replaced by the canonical momentum operators $ k^i\rightarrow\left(-i\partial^i+\frac{e}{\hbar}A^i \right)$, where $A^i$ is the vector potential. Hence $k^ik^j \rightarrow \left(-i\partial^i+\frac{e}{\hbar}A^i \right)\left(-i\partial^j+\frac{e}{\hbar}A^j \right)$, which can be decomposed into a gauge dependent symmetric component $\frac{1}{2}\left\{-i\partial^i+\frac{e}{\hbar}A^i,-i\partial^j+\frac{e}{\hbar}A^j\right\}$ and a gauge invariant anti-symmetric component $\frac{1}{2}\left[-i\partial^i+\frac{e}{\hbar}A^i,-i\partial^j+\frac{e}{\hbar}A^j\right]=-\frac{ie}{2\hbar}\sum_{k}\varepsilon_{ijk}B_k$ which contributes to orbital Zeeman's coupling. In summary, under magnetic field the total Hamiltonian consists of two parts: a gauge dependent part which leads to Landau level $H^{L}_{mm'}$ and a gauge invariant part which is the Zeeman's coupling $H^{Z}_{mm'}$, which can be written as,
\begin{align}
    \hat{H}^{L}_{mm'}=&\delta_{mm'}\epsilon_m+\bm{\hat{v}}_{mm'}\cdot\left(-i\bm{\nabla}+\frac{e}{\hbar}\bm{A}\right) \nonumber \\
    +&\sum_{ij}\hat{M}^{ij}_{mm'} \left(-i\partial^i+\frac{e}{\hbar}A^i \right)\left(-i\partial^j+\frac{e}{\hbar}A^j \right)
\end{align}
\begin{equation}\label{eq:gt}
   \hat{H}^{Z}_{mm'}=\mu_{B}(\bm{\hat{g}}^o_{mm'}+\bm{\hat{g}}^s_{mm'}) \cdot \bm{B}
\end{equation}
where 
\begin{align}
   \hat{M}^{ij}_{mm'}&=\delta_{mm'}\delta_{ij}\frac{\hbar^2}{2m_{\mathrm{e}}} + \frac{1}{4}\sum_{l}\left(\frac{1}{\epsilon_{m}-\epsilon_{l}}+\frac{1}{\epsilon_{m'}-\epsilon_{l}}\right) \nonumber \\
   & \times \left( \hat{v}^{i}_{ml} \hat{v}^{j}_{lm'}+\hat{v}^{j}_{ml} \hat{v}^{i}_{lm'} \right)
\end{align}
\begin{align}\label{eq:go}
   \bm{\hat{g}}^{o}_{mm'}=-\frac{i m_e}{2\hbar^2}\sum_{l,ijk}\left(\frac{1}{\epsilon_m-\epsilon_l}+\frac{1}{\epsilon_{m'}-\epsilon_{l}}\right)\hat{v}_{ml}^i\hat{v}_{lm'}^j\varepsilon_{ijk} \bm{e}_{k}
\end{align}
Here $\bm{e}_k$ is the unit direction vector.
The corresponding g-factor contributed by the spin part can be obtained straightforwardly as $\bm{\hat{g}}^{s}_{mm'}=\frac{2}{\hbar}\langle \psi_m | \bm{\hat{s}} | \psi_{m'} \rangle$, where $\bm{\hat{s}}$ is the spin operator.

\section{Vanishing Quantum Oscillations}

The first observable that manifests the momentum and field direction dependence of the Zeeman’s coupling is the ``spin-zero" effect\cite{shoenberg_magnetic_1984}, where the Shubnikov-de Haas Oscillation(SdH) or De Haas Von Alphen (dHV) effect vanishes when the field is applied along some special directions.
The SdH or dHV effect is the oscillation of the resistance or magnetic susceptibility that occurs under magnetic field. According to the Lifshitz-Kosevich formula, in the semiclassical limit the oscillations contributed by one FS are expressed as $\Delta\rho(B) \propto \sum_{ex} \cos(\hbar S_{\mathrm{ex}}/eB+\gamma+\phi)$, where $S_{ex}$ is the area of the extreme cross-section for the FS along the magnetic field, $\phi$ is the Berry phase over the boundary of the extreme cross-section $\partial S_{ex}$, the extra phase $\gamma$ equals $-\pi/4$ or $\pi/4$ for maximum or minimum extreme cross-section respectively and the sum is over the different extreme cross-section area.
A striking effect, called ``spin-zero" effect, manifests itself as the vanishing of quantum oscillations at some certain field directions, could happen when the doubly degenerate FS split into two under the Zeeman effect and thus contribute two oscillation terms with slightly different frequency and phases. 
More specifically, the Zeeman effect described by the g-factor tensor $\bm{\hat{g}}(\bm k)$ leads to not only the splitting of the cross-section area $S_{\mathrm{ex}}=S^0_{\mathrm{ex}}\pm\alpha B$ but also an extra splitting U(1) Berry phase $\pm\phi$ as we introduced above. Hence, by applying the sum-to-product identity, the total oscillations of the two splitting FS are expressed as
\begin{equation}\label{eq:deltarou}
    \Delta\rho(B) \propto \cos(\frac{\hbar S^0_{\mathrm{ex}}}{eB}+\gamma)\cos(\frac{\hbar \alpha}{e}+\phi)
\end{equation}
,where 
\begin{equation}\label{eq:alpha}
    \alpha=\oint_{\partial S^0_{\mathrm{e}x}}\frac{\sqrt{|\mathrm{det}[\mu_B \bm{\hat{g}}(\bm{k})\cdot \bm{e}_B]}|}{\chi|\nabla_{\bm{k}}\epsilon(\bm{k})\times\bm{e}_B|}\mathrm{d}{k}
\end{equation}
\begin{equation}\label{eq:berryphase}
    \phi=\oint_{\partial S^{0}_{ex}}\bm{A}_{+}(\bm{k})\cdot \mathrm{d}\bm{k}
\end{equation}
Here $\chi=-1$ for electron-like valley, $\chi=1$ for hole-like valley, $\epsilon(\bm{k})$ is the band energy of FS states at $\bm{k}$, $\phi$ is the Berry phase accumulated along one of the splitted FS's extreme area and $\bm{e}_B$ is the direction of the magnetic field. The last part of Eq.(\ref{eq:deltarou}) $R_s=\cos(\frac{\hbar \alpha}{e}+\phi)$ is the amplitude of oscillations which depends on the direction of magnetic field. The ``spin-zero" effect would happen at certain field directions where $R_s$ equals zero. 
We would emphasize that for materials with strong SOC in order to obtain the right field direction for the ``spin-zero" effect one has to compute both the coefficient $\alpha$ (for the splitting of FS area) and $\phi$ (for the splitting of U(1) Berry phase). 
As shown in Eq.(\ref{eq:alpha}) and (\ref{eq:berryphase}), both of them can be obtained from the k-dependent g-factor tensor on the FS (Details are given in Appendix \ref{sec:kp}). 

\begin{figure}[htbp] 
    \centering\includegraphics[width=3.4in,trim={0 0 2.9in 0},clip]{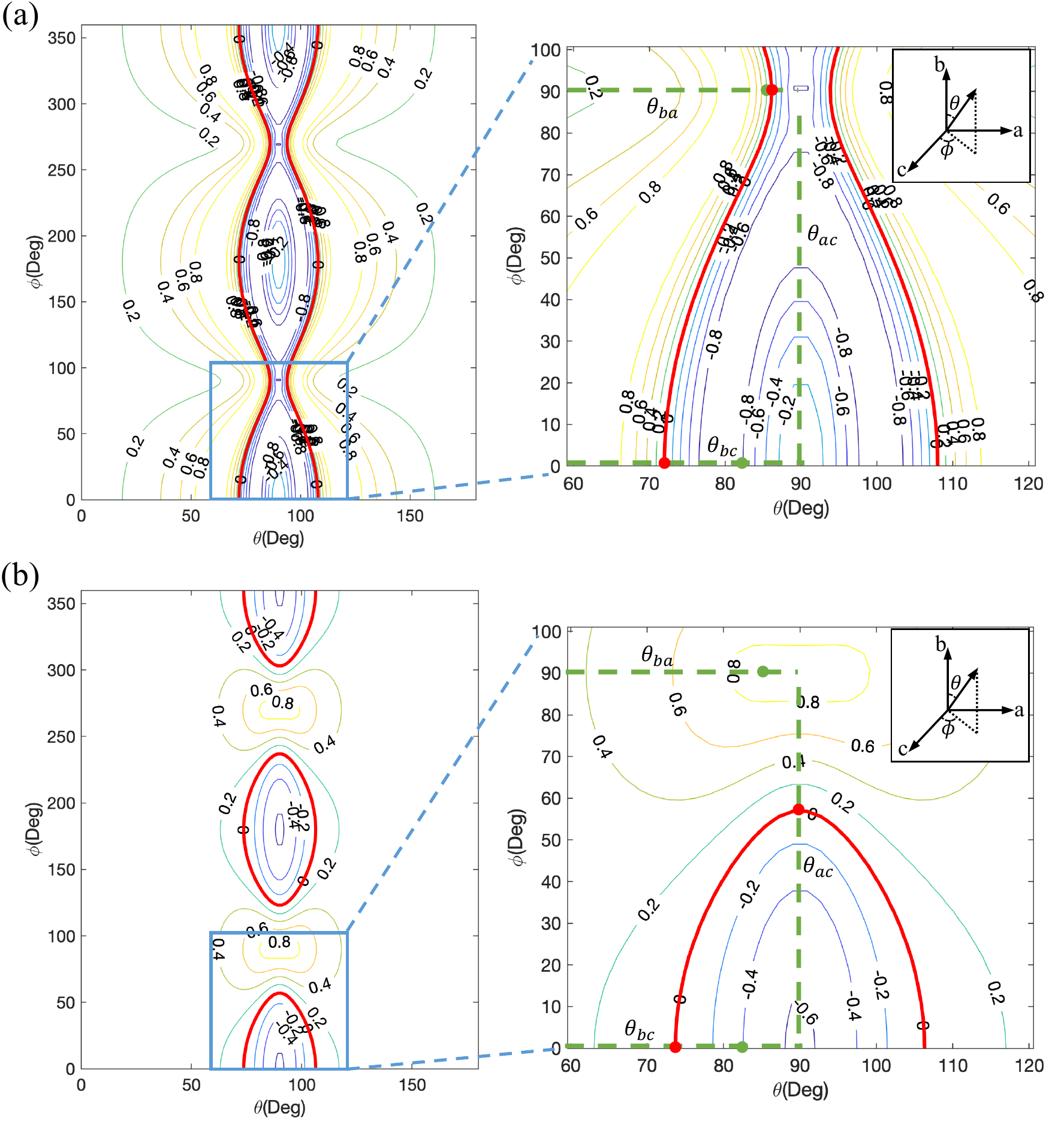} 
    \caption{\label{fig:vqo} Numerical calculations of the direction dependence of oscillation amplitudes factor $R_s(\bm{e}_B)$ based on GGA parameters given by Ref.\cite{song_first_principle} with Fermi energy $\mu_F=0.026$ eV($\mu_F=0$ eV is at the middle of the gap) in ZrTe$_5$. Oscillation amplitudes with Berry phase contribution (a) and without Berry phase contribution (b) are plotted individually. The angular for vanishing quantum oscillation are indicated by bold red line. And the angles measured in Ref.\cite{wang_vanishing_2018} are indicated by green dotted line. We only keep linear approximation of $\bm{k}$ because of the tiny FS in ZrTe5. Inset is a schematic illustration of the geometry for axes. The FS is an ellipsoid with principal semi-axes $k_a$=0.118 nm$^{-1}$,$k_b$=0.666 nm$^{-1}$ and $k_c$=0.153 nm$^{-1}$ which agrees well with the experiments\cite{wang_vanishing_2018}. }
\end{figure}

The narrow gap semiconductor ZrTe$_{5}$, which only has small ellipsoid FS around the $\Gamma$ point and shows strong anisotropy\cite{weng_transition-metal_2014,Liu2016,ZrTe5PhysRevB.93.115414,ZrTe5PhysRevB.31.7617}, is an ideal platform to study the ``spin-zero" effect. With the parameters given in Ref.\cite{song_first_principle}, we calculate the oscillation amplitude factor of the oscillation $R_s$ for all the field directions as illustrated in Fig.\ref{fig:vqo}, from which we can obtain the angles of ``spin-zero" which are summarized and compared with experiments\cite{wang_vanishing_2018} in Table.\ref{tab:vqo}. The theoretical results are consistent with the experimental results only when the Berry phase contributions are included. If we only consider the splitting of FS area, the corresponding results can't match the experimental data even qualitatively.

\begin{table}[htbp] 
    \caption{\label{tab:vqo} The angles of spin-zero effect}
    \begin{ruledtabular}
    \begin{tabular}{cccc}
    & \textrm{Theory} & \textrm{Theory(no Berry phases)} & \textrm{Experiment}\cite{wang_vanishing_2018}\\
    \colrule
    $\theta_{bc}$ & 72.0 & 73.7  & 83.8 \\
    $\theta_{ba}$ & 86.1 & $\varnothing$ & 86.5 \\
    $\theta_{ac}$ & $\varnothing$ & 56.8 & $\varnothing$  \\
    \end{tabular}
    \end{ruledtabular}
\end{table}

\section{In-Plane Hall effect and field induced topological metals}
\begin{figure}
    \centering\includegraphics[width=3.4in]{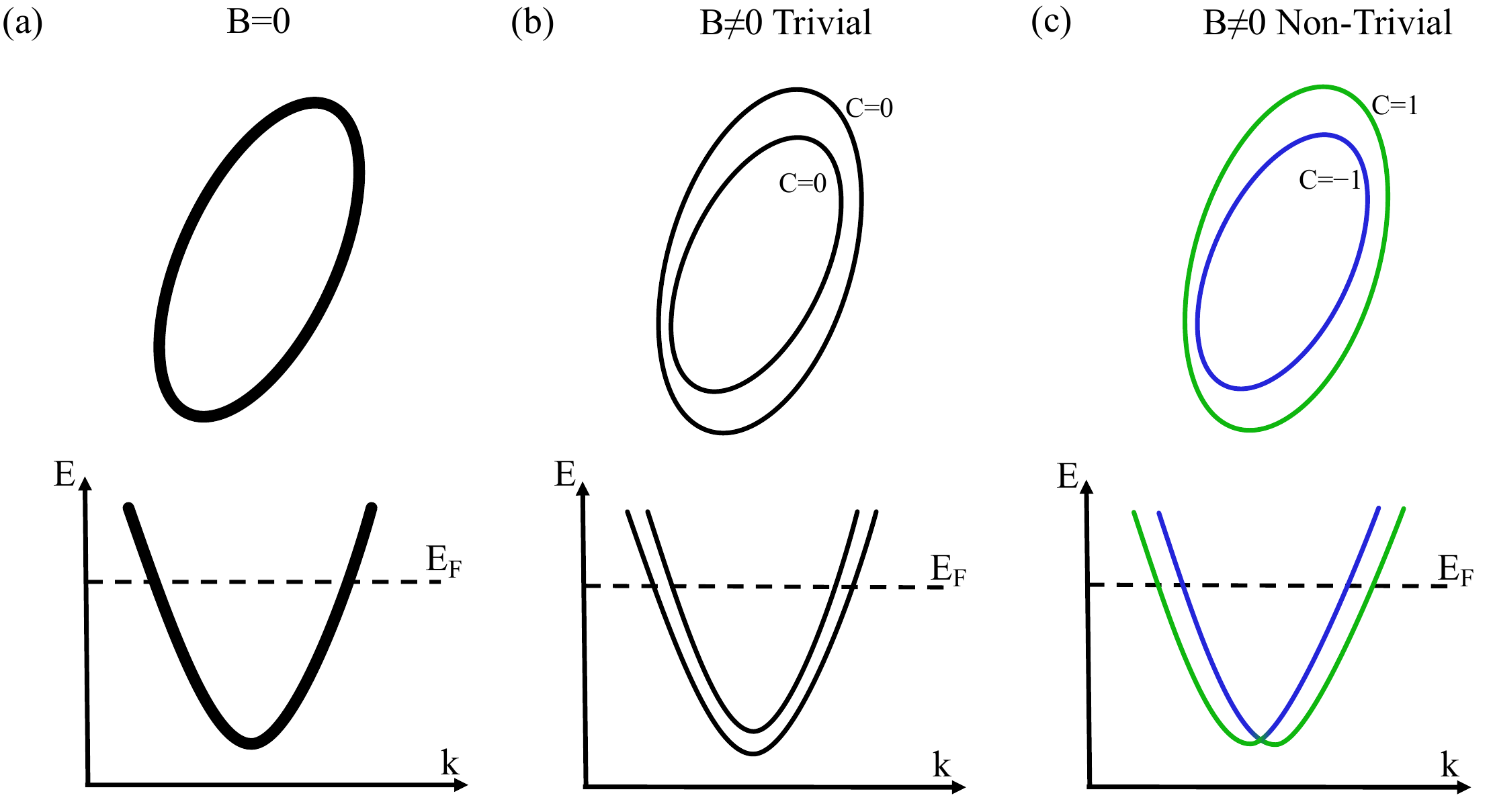} 
    \caption{\label{fig:schematic}Schematic diagrams for the mechanism of topological phase transition induced by the Zeeman’s coupling. The ellipses represent Fermi surfaces in a plane of momentum space.  When there is no magnetic field, the $\mathcal{PT}$ symmetry make the states doubly degenerate indicated by bold curve in (a). In the presence of magnetic field, the Zeeman’s coupling splits the degenerate states and Fermi surfaces split into two. For some directions of magnetic field, the Chern Number of both Fermi surfaces are zero as shown in (b) and there is no Weyl point enclosed by the Fermi surfaces. For other directions of magnetic field, the Fermi Surfaces have opposite non-zeros Chern Number as shown in (c) and there is a Weyl point enclosed by the Fermi surfaces. Here we only plot the most common and simple case. 
    }
\end{figure}

In the presence of magnetic field, the Zeeman effect splits the degenerate states into $\ket{\Psi_{\pm}(\bm{k})}$ with splitting energy $\pm\Delta \epsilon^z$. The Chern Numbers defined on each splitted Fermi surface are well-defined topological invariances that characterize the topological nature of that particular metal under magnetic field, which can be expressed as.
\begin{align}
    \mathcal{C}_{\pm}=\frac{1}{2\pi}\int_{FS}\mathrm{d}\bm{S}\cdot\bm{F}_{\pm}(\bm{k})
\end{align}
We can calculate the Chern Numbers of all the Fermi Surfaces with Willson loop method after including the Zeeman effect described by the g-factor tensor introduced previously.  
It is easy to prove that such FS Chern numbers will be only determined by the direction of the field and in general they can vary with the field direction, which defines the ``topological phase transition" on the FS. 
A schematic plot of Zeeman effect induced non-zero Chern Number on FS is shown in Fig.\ref{fig:schematic}. Theoretically, classifications of topological metals with $\mathcal{PT}$ symmetry has been studied in Ref.\cite{PTClassification}. Experimentally, the nonzero Chern numbers on FS will lead to similar chiral anomaly phenomena\cite{abj_1983,son_NMR_2013} and negative longitudinal magneto-resistance\cite{burkov_CA,burkov_NMR}, which has been observed already in some of these materials\cite{Na3Bi_NMR,PhysRevX.5.031023}.
 
Another way to manifest the Zeeman effect induced Berry phase and curvature on the FS is to look at the  anomalous Hall effect, which  is the Hall effect caused not by Lorentz force but the Zeeman induced Berry curvature around the FS. 
In order to minimize the interference from the Lorentz force, which exists for generic setup, it is better to apply the field within the plane of the experimental setup for the Hall measurement. Generally, anomalous Hall coefficient(AHC) can be expressed by the integral of Berry curvature over the Brillouin zone\cite{RevModPhys.82.1539,PhysRevB.74.195118} as
\begin{equation}
    \sigma_{ij}=-\varepsilon_{ijk}\frac{e^2}{\hbar}\sum_n \int_{BZ} \frac{\mathrm{d}^3\bm{k}}{(2\pi)^3}f\bm{(}\epsilon_n(\bm{k})\bm{)}F^k_n(\bm{k})
\end{equation}
Here, $F^k_n(\bm{k})$ represent the Berry curvature of the nth band in $k$ direction at $\bm{k}$ wave vector and $\varepsilon_{ijk}$ is the Levi-Civita notation.
At zero magnetic field, the two degenerate states with opposite Berry curvature will be always both occupied or unoccupied and their contribution will cancel each other. 
With the presence of magnetic field, the Zeeman effect will split these states and the net contribution to the AHC comes from a thin shell near the FS where only one of these otherwise degenerate states is occupied. 
Using the k-dependent g-factor tensor introduced above, we can express the AHC as $\sigma_{ij}=\lambda_{ij}(\bm{e}_B)B$, where
\begin{align}
    \lambda_{ij}(\bm{e}_B)=-\varepsilon_{ijk}\frac{e^3}{m_e}\int_{FS} \frac{\mathrm{d}{S}}{(2\pi)^3} \frac{\sqrt{|\mathrm{det}[\bm{\hat{g}}(\bm{k})\cdot \bm{e}_B]|}}{\chi|\nabla_{\bm{k}}\epsilon(\bm{k})|} F^k_{+}(\bm{k})
\end{align}
Here, $B$ is the strength of the magnetic field,  $F_{+}^k(\bm{k})$ represents the Berry curvature of splitted state $|\Psi_{+}(\bm{k})\rangle$ and the definition of $\chi$, $\bm{e}_B$ and $\epsilon(\bm{k})$ is same as Eq.(\ref{eq:alpha}). In particular, if the direction of magnetic field $\bm{e}_B$ is in the $ij$ plane, we will get the in-plane AHC, in which the voltage, current and magnetic field are all in the same plane.

\begin{figure}[htbp] 
    \centering\includegraphics[width=3.4in]{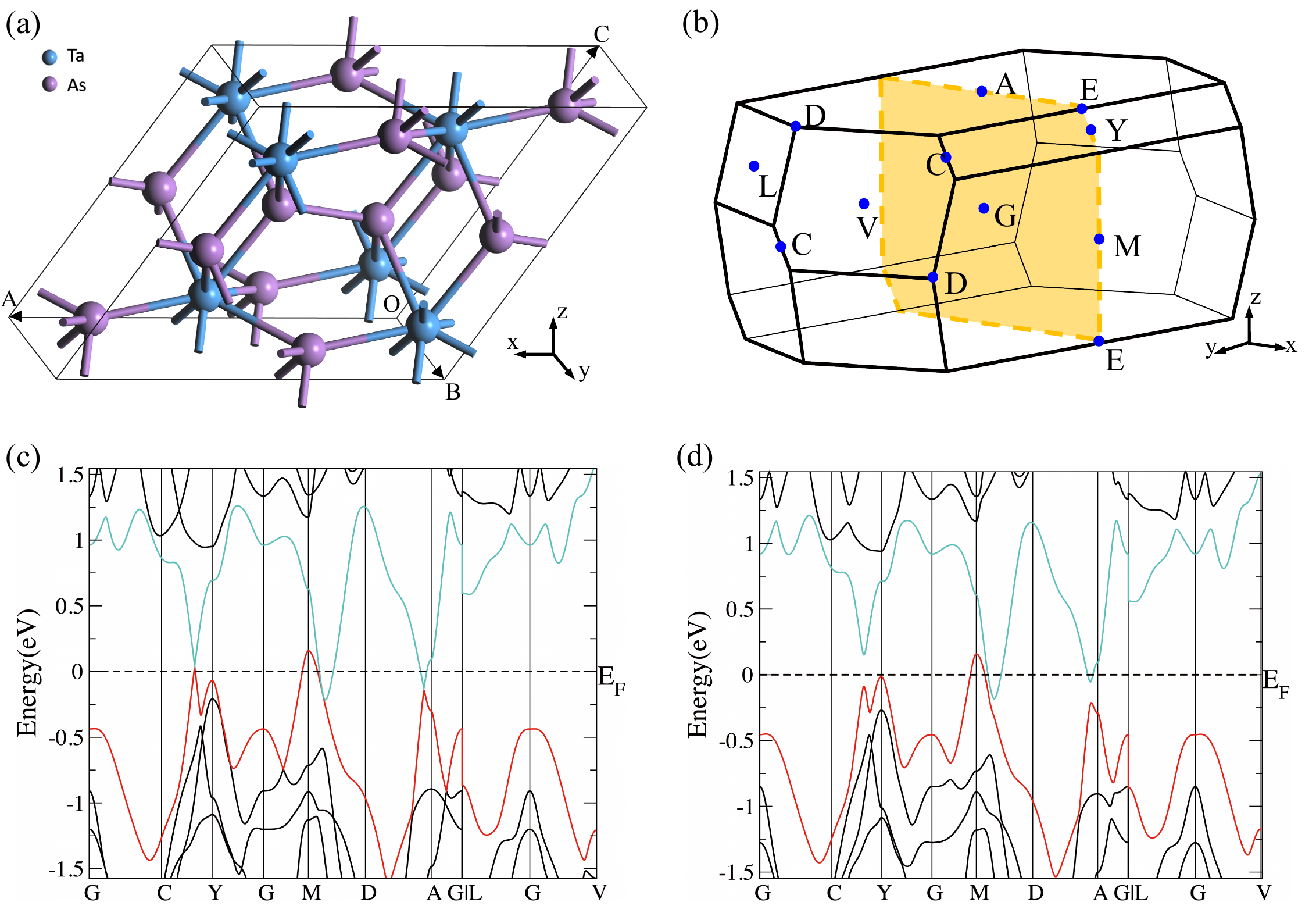} 
    \caption{\label{fig:crystal}Crystal structure, Brillouin zone and Band structure.  (a) Crystal structure of TaAs$_2$\cite{osti_1188980} (b) The first Brillouin zone for TaAs$_2$. The yellow plane shows the position of mirror plane and high symmetry points is indicated by blue dot. (c),(d) Band structure for TaAs$_2$ without SOC (c) and with SOC (d). From this we can find two electron-like pockets and one hole-like pocket.
    }
\end{figure}

In the present study, we take TaAs$_2$, an topological trivial semimetal with a number of tiny Fermi pockets, as a typical material example for the Zeeman effect induced FS Chern number and in-plane anomalous Hall effect. 
TaAs$_2$ crystallizes\cite{osti_1188980} in monoclinic structure with centrosymmetric space group of C2/m(No.12)  as shown in Fig.\ref{fig:crystal}(a). It has a binary axis (two-fold rotation symmetry) along y direction and a mirror plane perpendicular to y direction.
We performed the first principle calculations by using the generalized gradient approximation (GGA) for the exchange-correlation functional with the Vienna ab-initio simulation package(VASP). The cutoff energy for basis set is 400 eV and k-point sampling grid is $9 \times 9 \times 7$.

\begin{figure}[htbp] 
\centering\includegraphics[width=2.4in]{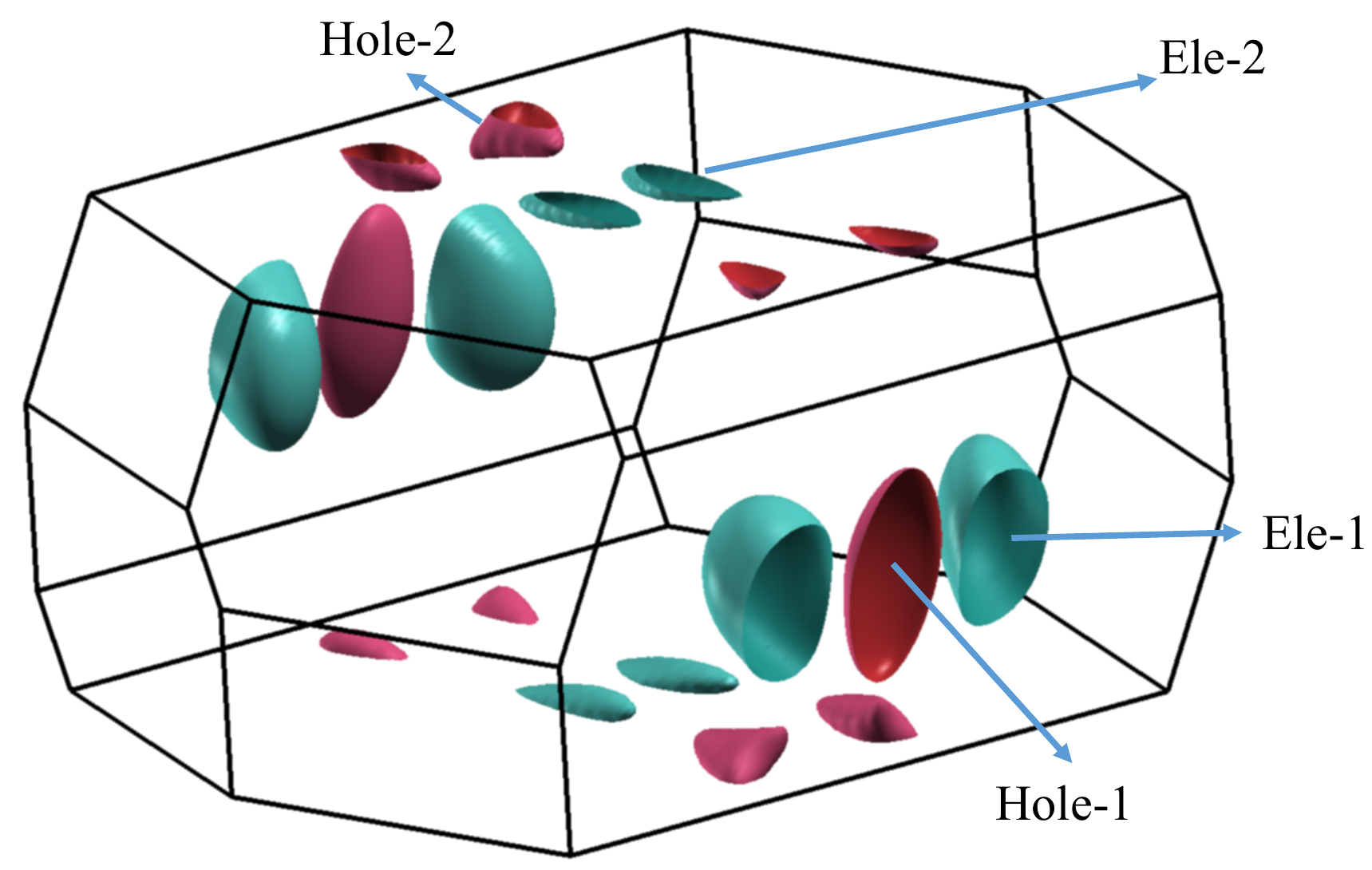}
\caption{\label{fig:FSs}Fermi Surfaces in the first Brillouin zone. According to the symmetry, there are 4 nonequivalent Fermi Surfaces in the first Brillouin zone. Two of them indicated in green are electron-like. The others indicated in red are hole-like.
} 
\end{figure}

The calculated electronic structure of TaAs$_2$ are shown in Fig.\ref{fig:crystal}(c)(d). Without SOC, the band structure of TaAs$_2$ contains a number of nodal lines, which is similar with TaAs. When the SOC is turned on, due to the presence of inversion symmetry, a complete gap will be opened on all the line nodes and the band structure has been checked to be completely trivial in contrast to its cousin TaAs.  Since the conduction and valence bands are still overlapping after including the SOC, TaAs$_2$ becomes a typical trivial semimetal with compensating electron and hole pockets. Around each pockets, the effective k $\cdot$p models are 4$\times$4 and can be generically described by a anisotropic Dirac equations with tiny mass terms (comparing to the chemical potential) leading to strong mixing between the conduction and valence bands near the band minimum(maximum), which is the microscopic origin for the strong k-dependent g-factor tensor.
The Fermi surface plot in Fig.\ref{fig:FSs} clearly shows that totally there are nine FS in the first Brillouin zone, which can be divided into four non-equivalent types according to the crystal symmetries. 
From the first principle results, we can construct the second order k$\cdot$p model Hamiltonian near the centers of each FS together with the k-dependent g-factor tensor, with which the FS Chern numbers after the Zeeman splitting have been calculated and shown in  Fig.\ref{fig:chern}. (Details of our calculations are given in Appendix \ref{sec:kp}) We find that indeed there are topological phase transitions in this material when we vary the direction of magnetic field. Please be noticed that the zero Chern Number of Hole-1 for all field directions is ensured by the inversion symmetry. For those FS with nonzero Chern numbers under the magnetic field, we have confirmed that there are Zeeman effect induced Weyl points enclosed within these FS. These Weyl points may contribute to negative magneto-resistance which has already been found in TaAs$_2$\cite{luo_anomalous_2016,yuan2016large}

\begin{figure}[htbp] 
    \centering\includegraphics[width=3.4in]{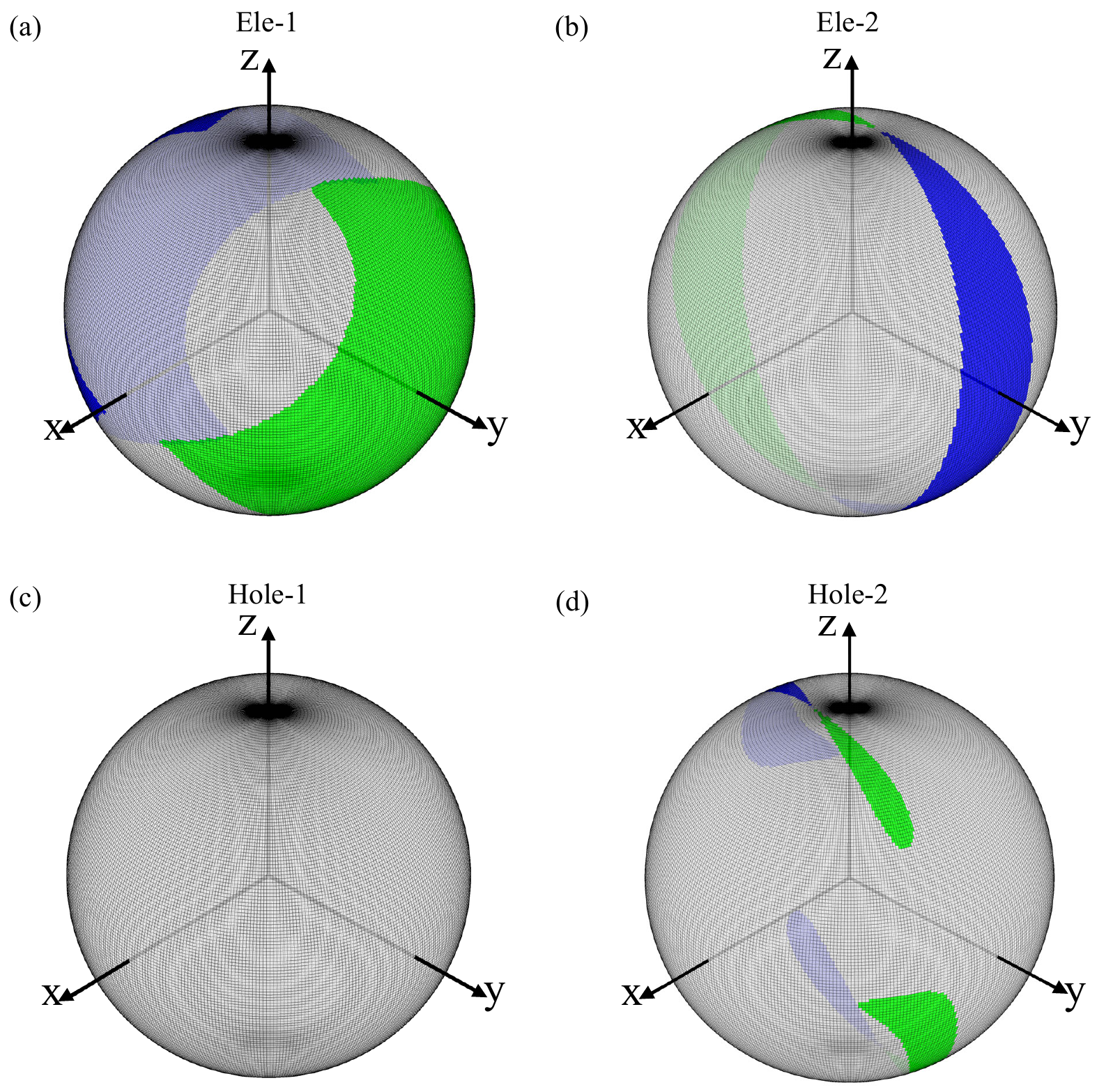}
    \caption{\label{fig:chern}The Chern Number of  4 non-equivalent Fermi Surface under different magnetic field directions. Directions of magnetic field are represented by points on the unit sphere. Chern Number=1 is indicated in green and Chern Number=$-1$ is indicated in blue. Here we plot the Chern Number of Fermi surface states $|\Psi_{+}(\bm{k})\rangle$. The other one $|\Psi_{-}(\bm{k})\rangle$ must has the opposite Chern Number. They are in accordance with the symmetry possessed by the Fermi surface. As a result of  inversion symmetry possessed by Hole-1, the Chern number is zero for all magnetic field directions shown in (c). 
    } 
\end{figure}

\begin{figure}[htbp] 
    \centering\includegraphics[width=2.4in]{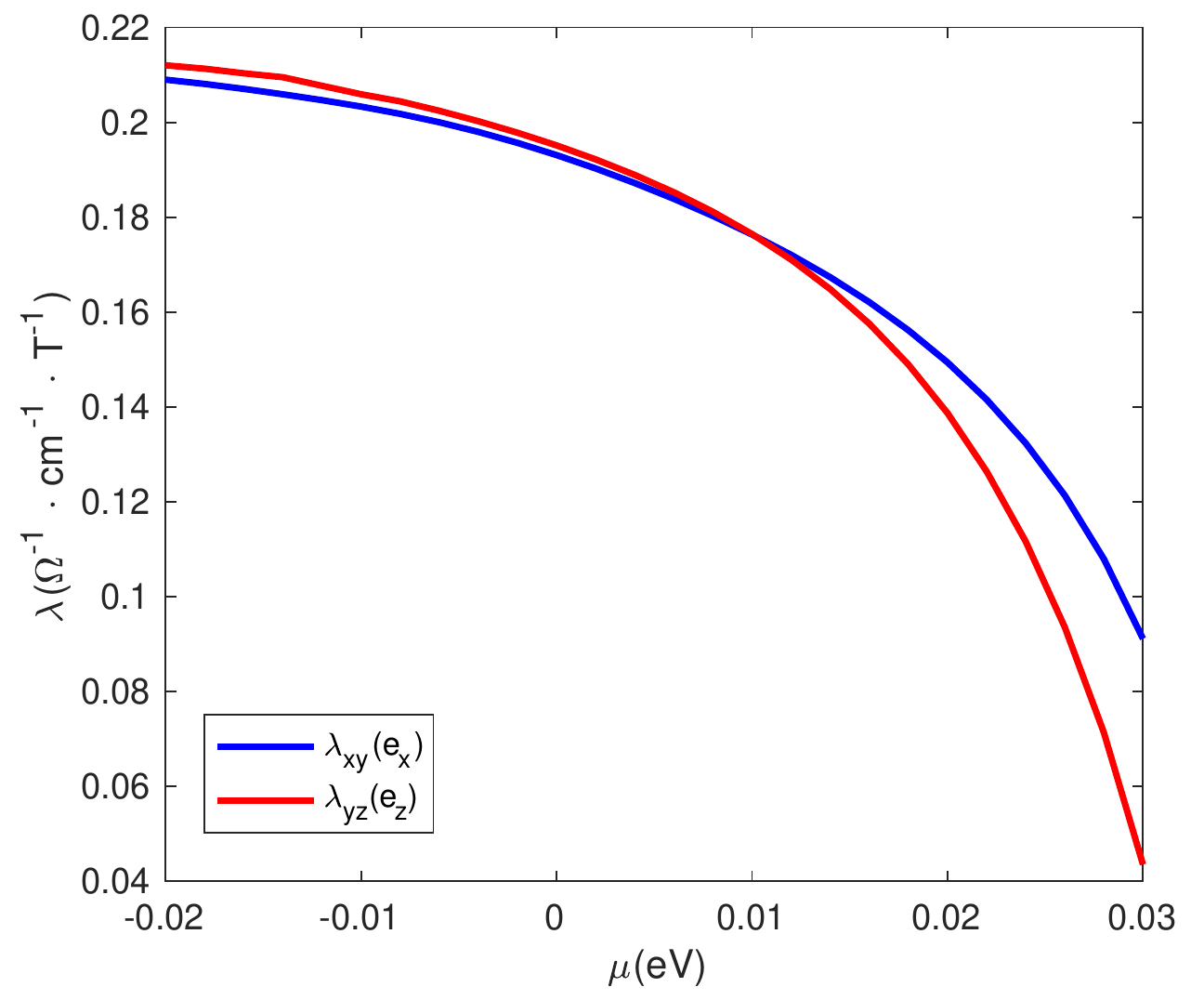}
    \caption{\label{fig:IPHE}In-plane anamolous Hall coefficient of TaAs$_2$ as the function of chemical potential. The In-plane anamolous Hall coefficient in xy plane $\lambda_{xy}(\hat{\bm{e}}_x)$ is indicated in blue line and coefficient in yz plane $\lambda_{yz}(\hat{\bm{e}}_z)$ is indicated in red line. 
    }
\end{figure}

We also calculated the in-plane anomalous Hall coefficients of TaAs$_2$ in xy, yz and zx planes.
By considering the crystal symmetries, the only allowed in-plane anomalous Hall coefficients are  $\lambda_{xy}(\hat{\bm{e}}_x)$ and $\lambda_{yz}(\hat{\bm{e}}_z)$, which are plotted in Fig.\ref{fig:IPHE} as the function of chemical potential. 
We find that significant magnitude of in-plane anomalous Hall effect can be realized in TaAs$_2$. 
Interestingly, the sign of such in-plane anomalous Hall coefficient keeps unchanged even when the carrier type changes from n to p as the function of chemical potential, which is qualitatively different with the ordinary Hall effect caused by Lorentz force.

\section{summary}
In summary, we have proposed that the Zeeman effect caused by the external magnetic field can be in general described by k-dependent g-factor tensor, which can be also viewed as the effective low energy Hamiltonian on the FS of central symmetric metals. The topological features hidden in such g-factor tensor can manifest themselves by generating additional Zeeman effect induced Berry's phase for Landau orbits, in-plane Hall effect and even nonzero Chern numbers on Zeeman splitted FS. All these exotic effects are demonstrated on two of the typical materials, ZrTe$_5$ and TaAs$_2$, by means of first principle calculations indicating that the effects proposed in the present study should widely exist in metals with inversion symmetry and strong SOC.

\section{acknowledgments}
We acknowledge the financial support from
the Hong Kong Research Grants Council (Project No.
GRF16300918), and thank professor Liyuan Zhang, Xiaosong Wu and Haizhou Lu for invaluable discussions.

\appendix
\section{\label{sec:PT}The $\mathcal{PT}$ symmetry}
Generally for a nonmagnetic centrosymmetric crystal, the representation of operation $\mathcal{P}$ and $\mathcal{T}$ can be written as following
\begin{align}
    &\mathcal{P}|\psi_{1}(\bm{k})\rangle=|\psi_{2}(\bm{-k})\rangle \qquad \mathcal{P}|\psi_{2}(\bm{k})\rangle=|\psi_{1}(\bm{-k})\rangle \\
    &\mathcal{T}|\psi_{1}(\bm{k})\rangle=|\psi_{2}(\bm{-k})\rangle \qquad \mathcal{T}|\psi_{2}(\bm{k})\rangle=-|\psi_{1}(\bm{-k})\rangle
\end{align} 
Hence for little group at any wave vector $\bm{k}$ there is at least one generator named $\mathcal{PT}$, and the representation is
\begin{align}
    \mathcal{PT}|\psi_{1}(\bm{k})\rangle=|\psi_{2}(\bm{k})\rangle \qquad \mathcal{PT}|\psi_{2}(\bm{k})\rangle=-|\psi_{1}(\bm{k})\rangle
\end{align}
or equivalently in matrix form
\begin{align}
    D(\mathcal{PT})=
    \begin{pmatrix}
        0 & -1 \\
        1 &  0 
    \end{pmatrix}
\end{align}
where $\mathcal{PT}|\psi_{i}(\bm{k})\rangle=|\psi_{j}(\bm{k})\rangle D_{ji}(\mathcal{PT})$. Under an unitary transformation $U\in \mathrm{U}(2)$, the representation of $D(\mathcal{PT})$ transforms as following 
\begin{align}\label{eq:reptrans}
    D'(\mathcal{PT})=U^{\dagger}D(\mathcal{PT})U^{*}
\end{align}  
where $|\psi'_{i}\rangle=|\psi_j\rangle U_{ji}$. Here we want to emphasize that the conjugate operation in Eq.(\ref{eq:reptrans}) comes from the antiunitary property\cite{sakurai} of time reversal operator $\mathcal{T}$.
Furthermore, we can divide the U(2) matrix $U$ into SU(2) part and $U(1)$ part as 
\begin{align} \label{eq:U}
    \hat{U}=\mathrm{e}^{i\theta} \exp \left({i \sum_{i} \frac{\sigma_i}{2} \omega_i} \right)
\end{align} 
By taking Eq.(\ref{eq:U}) into Eq.(\ref{eq:reptrans}) and applying anticommutation relation of Pauli matrix, we get
\begin{align}
    D'(\mathcal{PT})=e^{-2i\theta}D(PT)
\end{align}
Which means that the SU(2) part of the unitary transformation do not change the representation of $\mathcal{PT}$.
Generally because we want the representation of $\mathcal{PT}$ is independent of $\bm{k}$, $\theta$ should be $\bm{k}$ independent and $\omega$ could be $\bm{k}$ dependent.
Hence that is where the SU(2) gauge (SU(2) Berry connection and SU(2) Berry curvature) comes from. 

The Zeeman's coupling $\hat{H}^z(\bm{k})=\sum_{i}d_{i}(\bm{k})\sigma_i$  fix the SU(2) gauge $\hat{U}(\bm{k})$ which diagonalize the Zeeman's coupling
\begin{align}
    \hat{U}=
    \begin{pmatrix}
        \cos\frac{\theta_d}{2} & -e^{-i\phi_d}\sin\frac{\theta_d}{2} \\
        e^{i\phi_d}\sin\frac{\theta_d}{2} & \cos\frac{\theta_d}{2}
    \end{pmatrix}
\end{align}
where 
\begin{align}
    \cos{\theta_d}=\frac{d_z}{|\bm{d}|} \qquad e^{i\phi_d}=\frac{d_x+id_y}{\sqrt{d_x^2+d_y^2}}
\end{align}
By applying the antiunitary property and representation of $\mathcal{PT}$, we can prove that 
\begin{align}
    \bm{A}_1(\bm{k})&=i\langle\psi_1(\bm{k})|\partial_{\bm{k}}\psi_1(\bm{k})\rangle=i\langle\mathcal{PT}\partial_{\bm{k}}\psi_1(\bm{k})|\mathcal{PT}\psi_1(\bm{k})\rangle \nonumber \\
    &=i\langle\partial_{\bm{k}}\mathcal{PT}\psi_1(\bm{k})|\mathcal{PT}\psi_1(\bm{k})\rangle=i\langle\partial_{\bm{k}}\psi_2(\bm{k})|\psi_2(\bm{k})\rangle \nonumber \\
    &=-i\langle\psi_2(\bm{k})|\partial_{\bm{k}}\psi_2(\bm{k})\rangle \nonumber=-\bm{A}_2(\bm{k})
\end{align}
Hence the Berry curvature
\begin{align}
    \bm{F}_1(\bm{k})&=\bm{\nabla}_{\bm{k}} \times \bm{A}_1(\bm{k}) =-\bm{\nabla}_{\bm{k}} \times \bm{A}_2(\bm{k}) \nonumber \\
    &=-\bm{F}_2(\bm{k})
\end{align}

By saying the k$\cdot$p Hamiltonian satisfy the $\mathcal{PT}$ symmetry, we mean $\mathcal{PT}\mathcal{H}(\bm{k})=\mathcal{H}(\bm{k})\mathcal{PT}$  which in matrix form means that
\begin{align}\label{eq:dhd}
    D(\mathcal{PT})\hat{H}^{*}(\bm{k})=\hat{H}(\bm{k})D(\mathcal{PT})
\end{align}
Similarly the Zeeman's coupling satisfy the $\mathcal{PT}$ symmetry means that
\begin{align}
    D(\mathcal{PT})\bm{\hat{g}}^{*}(\bm{k})=-\bm{\hat{g}}(\bm{k})D(\mathcal{PT})
\end{align}
Here the minus sign comes from that under $\mathcal{PT}$ operation the magnetic field reverse the direction $\bm{B}\rightarrow-\bm{B}$

\section{\label{sec:kp}The k$\cdot$p Hamiltonian and 2$\times$2 g-factor tensor $\bm{\hat{g}}(\bm{k})$ calculations}

Now we discuss the first principle calculation of nonmagnetic centrosymmetric semimetals specifically. 
To balance the accuracy and the efficiency of the calculation we take the two-step down folding process as introduced below. 

As shown in Eq.(\ref{eq:go}), the $\bm{\hat{g}}(\bm{k})$ is inversely proportional to the energy difference ${\epsilon_m-\epsilon_l}$ between the low-energy subspace and the high-energy subspace. 
And as shown in Fig.\ref{fig:crystal}(d), for semimetals like TaAs$_2$, the FS contains a number of electron or hole pockets, which are called ``valleys" in this paper.
For each valley, we can choose the lowest conduction and highest valence bands as the low-energy subspace (all the rest bands as the high-energy subspace) for the first step and construct the 4$\times$4 k$\cdot$p model near the valley center $\bm{K}$. Since within each valley, the average energy distance between the high energy and low energy bands are much bigger than the band dispersion, we can safely approximate the g-factor tensor by a k independent constant for the 4$\times$4 model.

Here, in particular,  we take the representation of the $\mathcal{PT}$ as following
\begin{align}\label{eq:repofpt}
    D(\mathcal{PT})=
    \begin{pmatrix}
        0 & -i & 0  & 0 \\
        i & 0  & 0  & 0 \\
        0 & 0  & 0  & i \\
        0 & 0  & -i & 0 \\
    \end{pmatrix}
\end{align}
By applying Eq.(\ref{eq:dhd}), we find that the Hamiltonian has the following form
\begin{align}\label{eq:kphform}
    \hat{H}(\bm{k})=\Omega(\bm{k})+
    \begin{pmatrix}
        \Delta(\bm{k}) & \Lambda_{\mu}(\bm{k})\sigma_{\mu} \\
        \Lambda_{\mu}(\bm{k})\sigma_{\mu}^{\dagger} & -\Delta(\bm{k})
    \end{pmatrix}
\end{align}
where $\Omega(\bm{k})$, $\Delta(\bm{k})$ and $\Lambda_{\mu}(\bm{k})$ are \textit{real} functions of $\bm{k}$, $\mu=0,1,2,3$ and $\sigma_{\mu}$ are Pauli matrixes except $\sigma_0$ 
\begin{align}
    \sigma_0=i
    \begin{pmatrix}
        1 & 0 \\
        0 & 1
    \end{pmatrix}
\end{align}
And for $f=\Omega,\Delta,\Lambda_{\mu}$ the $\bm{k}$ dependence has the following form
\begin{align}
    f(\bm{k})=f^{(0)}+\sum_{i} f^{(1)}_i k_i +\sum_{ij}f^{(2)}_{ij}k_i k_j
\end{align} 
The corresponding 4$\times$4 g-factor matrix has the following form (Here we have included the spin contribution in the low-energy subspace and orbital contribution from the high-energy subspace)
\begin{align}
    \bm{\hat{g}}^{\left(4\right)}=
    \begin{pmatrix}
        \bm{g}^{a}             &  \bm{g}^{b}+i\bm{g}^{c} & \bm{g}^{e}+i\bm{g}^{f}    & \bm{g}^{g}+i\bm{g}^{h} \\
        \bm{g}^{b}-i\bm{g}^{c} &   -\bm{g}^{a}           & -\bm{g}^{g}+i\bm{g}^{h}   & \bm{g}^{e}-i\bm{g}^{f} \\
        \bm{g}^{e}-i\bm{g}^{f} & -\bm{g}^{g}-i\bm{g}^{h} & \bm{g}^{l}                & \bm{g}^{m}+i\bm{g}^{n} \\
        \bm{g}^{g}-i\bm{g}^{h} & \bm{g}^{e}+i\bm{g}^{f}  & \bm{g}^{m}-i\bm{g}^{n}    & -\bm{g}^{l}
    \end{pmatrix}
\end{align}
The corresponding parameters for ZrTe$_5$ are summarized in Ref.\cite{song_first_principle} and parameters for TaAs$_2$ are summarized in Table.\ref{tab:parameterskp} and Table.\ref{tab:parametersg}.
All these parameters are obtained by first principle calculation introduced in the main text and Ref.\cite{song_first_principle}.
\begin{table*}[htbp] 
    \caption{\label{tab:parameterskp} Parameters of k$\cdot$p Hamiltonian for TaAs$_{2}$}
    \begin{ruledtabular}
    \begin{tabular}{cddddcdddd}
    & \multicolumn{1}{c}{\textrm{Ele-1}} & \multicolumn{1}{c}{\textrm{Ele-2}} & \multicolumn{1}{c}{\textrm{Hole-1}} & \multicolumn{1}{c}{\textrm{Hole-2}} & & \multicolumn{1}{c}{\textrm{Ele-1}} & \multicolumn{1}{c}{\textrm{Ele-2}} & \multicolumn{1}{c}{\textrm{Hole-1}} & \multicolumn{1}{c}{\textrm{Hole-2}} \\
    \colrule
    $\Omega^{(0)}$                                & -0.27 & -0.15 & 0.37 & 0.19     &     $\Lambda^{(1)}_{2,3}$                         & 0.00 & 0.00 & 0.28 & -2.03      \\
    $\Omega^{(1)}_{1}$                            & -0.00 & -0.00 & -0.00 & -0.46   &     $\Lambda^{(1)}_{3,1}$                         & 0.26 & 0.78 & -0.00 & -0.64      \\
    $\Omega^{(1)}_{2}$                            & -0.74 & -0.86 & 0.00 & -0.80    &     $\Lambda^{(1)}_{3,2}$                         & -0.00 & -0.00 & -0.65 & -0.75      \\
    $\Omega^{(1)}_{3}$                            & 0.00 & -0.00 & -0.00 & -0.45    &     $\Lambda^{(1)}_{3,3}$                         & 0.36 & -3.33 & 0.00 & -1.02      \\
    $\Omega^{(2)}_{11}$                           & -3.31 & 3.39 & -0.41 & -3.52    &     $\Lambda^{(2)}_{0,11}$                        & -10.96 & -0.62 & -0.00 & 1.81      \\
    $\Omega^{(2)}_{12},\Omega^{(2)}_{21}$         & 0.00 & 0.00 & -0.00 & 1.55      &     $\Lambda^{(2)}_{0,12},\Lambda^{(2)}_{0,21}$   & 0.00 & 0.00 & -0.00 & -2.24      \\
    $\Omega^{(2)}_{13},\Omega^{(2)}_{31}$         & 7.21 & 0.79 & 1.26 & 2.67       &     $\Lambda^{(2)}_{0,13},\Lambda^{(2)}_{0,31}$   & 1.89 & -0.34 & 0.00 & 3.08      \\
    $\Omega^{(2)}_{22}$                           & 10.45 & -4.24 & -28.23 & -1.67  &     $\Lambda^{(2)}_{0,22}$                        & -8.07 & 2.59 & -0.00 & -1.55      \\
    $\Omega^{(2)}_{23},\Omega^{(2)}_{32}$         & 0.00 & -0.00 & 0.00 & 1.20      &     $\Lambda^{(2)}_{0,23},\Lambda^{(2)}_{0,32}$   & 0.00 & -0.00 & -0.00 & -0.58      \\
    $\Omega^{(2)}_{33}$                           & 4.06 & 3.56 & -0.18 & 6.32      &     $\Lambda^{(2)}_{0,33}$                        & -5.74 & -0.06 & -0.00 & 6.01      \\
    $\Delta^{(0)}$                                & -0.07 & -0.09 & -0.22 & -0.10   &     $\Lambda^{(2)}_{1,11}$                        & -0.00 & -0.00 & -0.00 & -1.35      \\
    $\Delta^{(1)}_{1}$                            & -0.00 & 0.00 & -0.00 & 0.46     &     $\Lambda^{(2)}_{1,12},\Lambda^{(2)}_{1,21}$   & 7.81 & 1.02 & -0.00 & 0.40      \\
    $\Delta^{(1)}_{2}$                            & -0.74 & -0.88 & -0.00 & 0.80    &     $\Lambda^{(2)}_{1,13},\Lambda^{(2)}_{1,31}$   & 0.00 & 0.00 & -0.00 & -1.31      \\
    $\Delta^{(1)}_{3}$                            & 0.00 & -0.00 & -0.00 & 0.45     &     $\Lambda^{(2)}_{1,22}$                        & -0.00 & -0.00 & 0.00 & -1.69      \\
    $\Delta^{(2)}_{11}$                           & -4.13 & 0.46 & 0.56 & 0.14      &     $\Lambda^{(2)}_{1,23},\Lambda^{(2)}_{1,32}$   & 2.33 & -0.93 & -0.00 & -0.44      \\
    $\Delta^{(2)}_{12},\Delta^{(2)}_{21}$         & 0.00 & -0.00 & -0.00 & 1.43     &     $\Lambda^{(2)}_{1,33}$                        & 0.00 & -0.00 & -0.00 & -3.71      \\
    $\Delta^{(2)}_{13},\Delta^{(2)}_{31}$         & 0.30 & -0.03 & 5.10 & -0.26     &     $\Lambda^{(2)}_{2,11}$                        & -3.67 & -1.52 & -0.00 & -0.08      \\
    $\Delta^{(2)}_{22}$                           & -3.05 & 0.29 & 21.01 & 0.77     &     $\Lambda^{(2)}_{2,12},\Lambda^{(2)}_{2,21}$   & 0.00 & -0.00 & 0.00 & -2.53      \\
    $\Delta^{(2)}_{23},\Delta^{(2)}_{32}$         & -0.00 & -0.00 & 0.00 & 0.42     &     $\Lambda^{(2)}_{2,13},\Lambda^{(2)}_{2,31}$   & 1.21 & -1.26 & 0.00 & -0.71      \\
    $\Delta^{(2)}_{33}$                           & -2.45 & -0.04 & -1.22 & -2.29   &     $\Lambda^{(2)}_{2,22}$                        & -1.23 & 7.13 & -0.00 & -2.15      \\
    $\Lambda^{(1)}_{0,1}$                         & -0.00 & -0.00 & -3.27 & -0.44   &     $\Lambda^{(2)}_{2,23},\Lambda^{(2)}_{2,32}$   & 0.00 & 0.00 & 0.00 & -1.85      \\
    $\Lambda^{(1)}_{0,2}$                         & -1.22 & 0.83 & 0.00 & -1.94     &     $\Lambda^{(2)}_{2,33}$                        & 0.47 & -1.15 & -0.00 & 1.32      \\
    $\Lambda^{(1)}_{0,3}$                         & 0.00 & -0.00 & -0.85 & 1.88     &     $\Lambda^{(2)}_{2,11}$                        & 0.00 & 0.00 & -0.00 & 1.32      \\
    $\Lambda^{(1)}_{1,1}$                         & 0.35 & -0.49 & -0.00 & -0.14    &     $\Lambda^{(2)}_{3,12},\Lambda^{(2)}_{3,21}$   & 1.32 & -1.94 & 0.00 & -1.56      \\
    $\Lambda^{(1)}_{1,2}$                         & 0.00 & 0.00 & -0.06 & 0.28      &     $\Lambda^{(2)}_{3,13},\Lambda^{(2)}_{3,31}$   & -0.00 & -0.00 & 0.00 & 0.20      \\
    $\Lambda^{(1)}_{1,3}$                         & 0.14 & 2.17 & 0.00 & -0.73      &     $\Lambda^{(2)}_{3,22}$                        & -0.00 & -0.00 & -0.00 & -0.70      \\
    $\Lambda^{(1)}_{2,1}$                         & 0.00 & -0.00 & 0.03 & -0.79     &     $\Lambda^{(2)}_{3,23},\Lambda^{(2)}_{3,32}$   & 0.73 & 1.70 & 0.00 & -1.17      \\
    $\Lambda^{(1)}_{2,2}$                         & 0.47 & 2.81 & -0.00 & -1.00     &     $\Lambda^{(2)}_{3,33}$                        & 0.00 & 0.00 & 0.00 & 1.59      \\
    \end{tabular}
    \end{ruledtabular}
\end{table*}

\begin{table*}[htbp] 
    \caption{\label{tab:parametersg} Parameters for $4\times 4$ g-factor for TaAs$_2$}
    \begin{ruledtabular}
    \begin{tabular}{cddddcdddd}
    & \multicolumn{1}{c}{\textrm{Ele-1}} & \multicolumn{1}{c}{\textrm{Ele-2}} & \multicolumn{1}{c}{\textrm{Hole-1}} & \multicolumn{1}{c}{\textrm{Hole-2}} & & \multicolumn{1}{c}{\textrm{Ele-1}} & \multicolumn{1}{c}{\textrm{Ele-2}} & \multicolumn{1}{c}{\textrm{Hole-1}} & \multicolumn{1}{c}{\textrm{Hole-2}} \\
    \colrule
    $g^a_1$ & -0.71 & -0.28 & -0.60 & 0.28 &     $g^g_1$ & 0.00 & -0.00 & -0.00 & 0.10 \\
    $g^a_2$ & -0.00 & 0.00 & -0.00 & -0.08 &     $g^g_2$ & -1.50 & -0.24 & 0.00 & -0.41 \\
    $g^a_3$ & -0.64 & 0.44 & -0.88 & 0.12  &     $g^g_3$ & 0.00 & -0.00 & -0.00 & -0.44 \\ 
    $g^b_1$ & 0.06 & -0.07 & 0.46 & 0.03   &     $g^h_1$ & 0.42 & -0.17 & -0.00 & 0.06 \\  
    $g^b_2$ & 0.00 & -0.00 & 0.00 & 0.12   &     $g^h_2$ & -0.00 & 0.00 & -0.00 & -0.74 \\ 
    $g^b_3$ & -1.63 & 0.45 & -0.63 & -0.44 &     $g^h_3$ & 0.70 & -0.93 & -0.00 & 0.63 \\  
    $g^c_1$ & 0.00 & 0.00 & -0.00 & -0.93  &     $g^l_1$ & 0.25 & 0.64 & 0.21 & 0.44 \\  
    $g^c_2$ & -0.25 & -0.17 & 0.97 & 0.76  &     $g^l_2$ & 0.00 & -0.00 & -0.00 & 0.78 \\  
    $g^c_3$ & 0.00 & -0.00 & 0.00 & 0.70   &     $g^l_3$ & -0.59 & -0.15 & -2.06 & 0.93 \\ 
    $g^e_1$ & 0.00 & 0.00 & 0.00 & -0.02   &     $g^m_1$ & -0.54 & 1.82 & 0.72 & -0.05 \\ 
    $g^e_2$ & 0.04 & -0.92 & -0.00 & -0.17 &     $g^m_2$ & -0.00 & 0.00 & 0.00 & -0.10 \\  
    $g^e_3$ & 0.00 & 0.00 & 0.00 & -0.39   &     $g^m_3$ & -1.37 & 0.42 & -1.43 & 0.03 \\ 
    $g^f_1$ & -0.81 & 0.18 & -0.00 & -0.07 &     $g^n_1$ & -0.00 & 0.00 & -0.00 & -0.41 \\   
    $g^f_2$ & -0.00 & 0.00 & 0.00 & 0.53   &     $g^n_2$ & -0.75 & 0.17 & 0.91 & -0.86 \\ 
    $g^f_3$ & 0.08 & -0.45 & -0.00 & 0.60  &     $g^n_3$ & -0.00 & -0.00 & -0.00 & -0.30\\  

    \end{tabular}
    \end{ruledtabular}
\end{table*}
The eigenvalues of k$\cdot$p Hamiltonian is
\begin{align}
    \epsilon^{\pm}(\bm{k})=\Omega(\bm{k}) \pm \sqrt{\Delta(\bm{k})^2+\sum_{\mu} \Lambda_{\mu}(\bm{k})^2}
\end{align}
And the unitary transformation which can diagonalize the k$\cdot$p Hamiltonian is
\begin{widetext}
\begin{align}
    \hat{V}(\bm{k})=
    \begin{pmatrix}
        \sqrt{\frac{\epsilon^{-}-\Omega+\Delta}{2(\epsilon^{-}-\Omega)}}
        \begin{pmatrix}
            \begin{pmatrix} 
                1 & 0 \\ 0 & 1 
            \end{pmatrix} \\ 
            \frac{1}{\epsilon^{-}-\Omega+\Delta}
            \begin{pmatrix} 
                \Lambda_3-\mathrm{i}\Lambda_0 & \Lambda_1-\mathrm{i}\Lambda_2 \\
                \Lambda_1+\mathrm{i}\Lambda_2 &-\Lambda_3-\mathrm{i}\Lambda_0 
            \end{pmatrix}
        \end{pmatrix}
        \sqrt{\frac{\epsilon^{+}-\Omega-\Delta}{2(\epsilon^{+}-\Omega)}}
        \begin{pmatrix}
            \frac{1}{\epsilon^{+}-\Omega-\Delta}
            \begin{pmatrix}
                \Lambda_3+\mathrm{i}\Lambda_0 & \Lambda_1-\mathrm{i}\Lambda_2 \\
                \Lambda_1+\mathrm{i}\Lambda_2 & -\Lambda_3+\mathrm{i}\Lambda_0 
            \end{pmatrix} \\
            \begin{pmatrix} 
                1 & 0 \\ 0 & 1 
            \end{pmatrix}
        \end{pmatrix}
    \end{pmatrix}
\end{align}
\end{widetext}
Now we take the second down folding process. Around any wave vector $\bm{k}$, the k$\cdot$p Hamiltonian in linear approximation transformed by the unitary transformation $\hat{V}(\bm{k})$ at $\bm{k}$ has the following form 
\begin{align}
    \tilde{H}_{mm'}(\bm{k}+\Delta\bm{k})&=\hat{V}(\bm{k})^{\dagger}H(\bm{k}+\Delta\bm{k})\hat{V}(\bm{k})\\
    &=\epsilon_{m}(\bm{k})\delta_{mm'}+\tilde{\bm{v}}_{mm'}(\bm{k})\cdot\Delta\bm{k}
\end{align}
where $\tilde{v}(\bm{k})=\hat{V}^{\dagger}(\bm{k})\frac{\partial \hat{H}(\bm{k})}{\partial \bm{k}}\hat{V}(\bm{k})$, $\epsilon_{1,2}=\epsilon_{-}$(named hole-like bands), $\epsilon_{3,4}=\epsilon_{+}$(named electron-like bands). 
Then the g-factor for hole-like (electron-like) bands contributed by the electron-like (hole-like) bands can be calculated with Eq.(\ref{eq:go}) as following which is $\bm{k}$ dependent
\begin{align}
    \bm{\hat{g}}^{\left(2\right)}_{pp'}(\bm{k})=\frac{im_{\mathrm{e}}}{\hbar^2}\frac{\chi}{\epsilon^{+}-\epsilon^{-}}\sum_{q,ijk}\tilde{v}_{pq,i}(\bm{k})\tilde{v}_{qp',j}(\bm{k})\varepsilon_{ijk}\bm{e}_k
\end{align}
where, for electron-like bands, $\chi=-1$, $p,p'=3,4$ and $q=1,2$; for hole-like bands, $\chi=1$, $p,p'=1,2$ and $q=3,4$.
Hence the total g-factor is 
\begin{align}
    \bm{\hat{g}}_{pp'}(\bm{k})=\bm{\hat{g}}^{\left(2\right)}_{pp'}(\bm{k})+\sum_{mm'}\hat{V}^{\dagger}_{pm}\left(\bm{k}\right)\bm{\hat{g}}^{\left(4\right)}_{mm'}V_{m'p'}\left(\bm{k}\right)
\end{align}

\bibliography{Zeeman}

\begin{thebibliography}{43}%
\makeatletter
\providecommand \@ifxundefined [1]{%
 \@ifx{#1\undefined}
}%
\providecommand \@ifnum [1]{%
 \ifnum #1\expandafter \@firstoftwo
 \else \expandafter \@secondoftwo
 \fi
}%
\providecommand \@ifx [1]{%
 \ifx #1\expandafter \@firstoftwo
 \else \expandafter \@secondoftwo
 \fi
}%
\providecommand \natexlab [1]{#1}%
\providecommand \enquote  [1]{``#1''}%
\providecommand \bibnamefont  [1]{#1}%
\providecommand \bibfnamefont [1]{#1}%
\providecommand \citenamefont [1]{#1}%
\providecommand \href@noop [0]{\@secondoftwo}%
\providecommand \href [0]{\begingroup \@sanitize@url \@href}%
\providecommand \@href[1]{\@@startlink{#1}\@@href}%
\providecommand \@@href[1]{\endgroup#1\@@endlink}%
\providecommand \@sanitize@url [0]{\catcode `\\12\catcode `\$12\catcode
  `\&12\catcode `\#12\catcode `\^12\catcode `\_12\catcode `\%12\relax}%
\providecommand \@@startlink[1]{}%
\providecommand \@@endlink[0]{}%
\providecommand \url  [0]{\begingroup\@sanitize@url \@url }%
\providecommand \@url [1]{\endgroup\@href {#1}{\urlprefix }}%
\providecommand \urlprefix  [0]{URL }%
\providecommand \Eprint [0]{\href }%
\providecommand \doibase [0]{https://doi.org/}%
\providecommand \selectlanguage [0]{\@gobble}%
\providecommand \bibinfo  [0]{\@secondoftwo}%
\providecommand \bibfield  [0]{\@secondoftwo}%
\providecommand \translation [1]{[#1]}%
\providecommand \BibitemOpen [0]{}%
\providecommand \bibitemStop [0]{}%
\providecommand \bibitemNoStop [0]{.\EOS\space}%
\providecommand \EOS [0]{\spacefactor3000\relax}%
\providecommand \BibitemShut  [1]{\csname bibitem#1\endcsname}%
\let\auto@bib@innerbib\@empty
\bibitem [{\citenamefont {Cohen}\ and\ \citenamefont
  {Blount}(1960)}]{cohen_g-factor_1960}%
  \BibitemOpen
  \bibfield  {author} {\bibinfo {author} {\bibfnamefont {M.~H.}\ \bibnamefont
  {Cohen}}\ and\ \bibinfo {author} {\bibfnamefont {E.~I.}\ \bibnamefont
  {Blount}},\ }\bibfield  {title} {\bibinfo {title} {The g-factor and de
  haas-van alphen effect of electrons in bismuth},\ }\href
  {https://doi.org/10.1080/14786436008243294} {\bibfield  {journal} {\bibinfo
  {journal} {Philosophical Magazine}\ }\textbf {\bibinfo {volume} {5}},\
  \bibinfo {pages} {115} (\bibinfo {year} {1960})}\BibitemShut {NoStop}%
\bibitem [{\citenamefont {Luttinger}\ and\ \citenamefont
  {Kohn}(1955)}]{luttinger_perturbation}%
  \BibitemOpen
  \bibfield  {author} {\bibinfo {author} {\bibfnamefont {J.~M.}\ \bibnamefont
  {Luttinger}}\ and\ \bibinfo {author} {\bibfnamefont {W.}~\bibnamefont
  {Kohn}},\ }\bibfield  {title} {\bibinfo {title} {Motion of {Electrons} and
  {Holes} in {Perturbed} {Periodic} {Fields}},\ }\href
  {https://doi.org/10.1103/PhysRev.97.869} {\bibfield  {journal} {\bibinfo
  {journal} {Physical Review}\ }\textbf {\bibinfo {volume} {97}},\ \bibinfo
  {pages} {869} (\bibinfo {year} {1955})}\BibitemShut {NoStop}%
\bibitem [{\citenamefont {Song}\ \emph {et~al.}(2015)\citenamefont {Song},
  \citenamefont {Sun}, \citenamefont {Xu}, \citenamefont {Nie}, \citenamefont
  {Weng}, \citenamefont {Fang},\ and\ \citenamefont
  {Dai}}]{song_first_principle}%
  \BibitemOpen
  \bibfield  {author} {\bibinfo {author} {\bibfnamefont {Z.-D.}\ \bibnamefont
  {Song}}, \bibinfo {author} {\bibfnamefont {S.}~\bibnamefont {Sun}}, \bibinfo
  {author} {\bibfnamefont {Y.-F.}\ \bibnamefont {Xu}}, \bibinfo {author}
  {\bibfnamefont {S.-M.}\ \bibnamefont {Nie}}, \bibinfo {author} {\bibfnamefont
  {H.-M.}\ \bibnamefont {Weng}}, \bibinfo {author} {\bibfnamefont
  {Z.}~\bibnamefont {Fang}},\ and\ \bibinfo {author} {\bibfnamefont
  {X.}~\bibnamefont {Dai}},\ }\href@noop {} {\bibinfo {title} {First principle
  calculation of the effective zeeman's couplings in topological materials}}
  (\bibinfo {year} {2015}),\ \Eprint {https://arxiv.org/abs/1512.05084}
  {arXiv:1512.05084 [cond-mat.mtrl-sci]} \BibitemShut {NoStop}%
\bibitem [{\citenamefont {Mikitik}\ and\ \citenamefont
  {Sharlai}(1999)}]{mikitik1999manifestation}%
  \BibitemOpen
  \bibfield  {author} {\bibinfo {author} {\bibfnamefont {G.~P.}\ \bibnamefont
  {Mikitik}}\ and\ \bibinfo {author} {\bibfnamefont {Y.~V.}\ \bibnamefont
  {Sharlai}},\ }\bibfield  {title} {\bibinfo {title} {Manifestation of berry's
  phase in metal physics},\ }\href
  {https://doi.org/10.1103/PhysRevLett.82.2147} {\bibfield  {journal} {\bibinfo
   {journal} {Phys. Rev. Lett.}\ }\textbf {\bibinfo {volume} {82}},\ \bibinfo
  {pages} {2147} (\bibinfo {year} {1999})}\BibitemShut {NoStop}%
\bibitem [{\citenamefont {Zhang}\ \emph {et~al.}(2005)\citenamefont {Zhang},
  \citenamefont {Tan}, \citenamefont {Stormer},\ and\ \citenamefont
  {Kim}}]{zhang2005experimental}%
  \BibitemOpen
  \bibfield  {author} {\bibinfo {author} {\bibfnamefont {Y.}~\bibnamefont
  {Zhang}}, \bibinfo {author} {\bibfnamefont {Y.-W.}\ \bibnamefont {Tan}},
  \bibinfo {author} {\bibfnamefont {H.~L.}\ \bibnamefont {Stormer}},\ and\
  \bibinfo {author} {\bibfnamefont {P.}~\bibnamefont {Kim}},\ }\bibfield
  {title} {\bibinfo {title} {Experimental observation of the quantum hall
  effect and berry's phase in graphene},\ }\href
  {https://doi.org/10.1038/nature04235} {\bibfield  {journal} {\bibinfo
  {journal} {Nature}\ }\textbf {\bibinfo {volume} {438}},\ \bibinfo {pages}
  {201} (\bibinfo {year} {2005})}\BibitemShut {NoStop}%
\bibitem [{\citenamefont {Novoselov}\ \emph {et~al.}(2006)\citenamefont
  {Novoselov}, \citenamefont {McCann}, \citenamefont {Morozov}, \citenamefont
  {Fal'ko}, \citenamefont {Katsnelson}, \citenamefont {Zeitler}, \citenamefont
  {Jiang}, \citenamefont {Schedin},\ and\ \citenamefont
  {Geim}}]{Novoselov2006}%
  \BibitemOpen
  \bibfield  {author} {\bibinfo {author} {\bibfnamefont {K.~S.}\ \bibnamefont
  {Novoselov}}, \bibinfo {author} {\bibfnamefont {E.}~\bibnamefont {McCann}},
  \bibinfo {author} {\bibfnamefont {S.~V.}\ \bibnamefont {Morozov}}, \bibinfo
  {author} {\bibfnamefont {V.~I.}\ \bibnamefont {Fal'ko}}, \bibinfo {author}
  {\bibfnamefont {M.~I.}\ \bibnamefont {Katsnelson}}, \bibinfo {author}
  {\bibfnamefont {U.}~\bibnamefont {Zeitler}}, \bibinfo {author} {\bibfnamefont
  {D.}~\bibnamefont {Jiang}}, \bibinfo {author} {\bibfnamefont
  {F.}~\bibnamefont {Schedin}},\ and\ \bibinfo {author} {\bibfnamefont {A.~K.}\
  \bibnamefont {Geim}},\ }\bibfield  {title} {\bibinfo {title} {Unconventional
  quantum hall effect and berry's phase of 2p in bilayer graphene},\ }\href
  {https://doi.org/10.1038/nphys245} {\bibfield  {journal} {\bibinfo  {journal}
  {Nature Physics}\ }\textbf {\bibinfo {volume} {2}},\ \bibinfo {pages} {177}
  (\bibinfo {year} {2006})}\BibitemShut {NoStop}%
\bibitem [{\citenamefont {Shoenberg}()}]{shoenberg_magnetic_1984}%
  \BibitemOpen
  \bibfield  {author} {\bibinfo {author} {\bibfnamefont {D.}~\bibnamefont
  {Shoenberg}},\ }\href@noop {} {\emph {\bibinfo {title} {Magnetic oscillations
  in metals}}},\ Cambridge monographs on physics\ (\bibinfo  {publisher}
  {Cambridge University Press})\ p.\ \bibinfo {pages} {448}\BibitemShut
  {NoStop}%
\bibitem [{\citenamefont {Nagaosa}\ \emph {et~al.}(2010)\citenamefont
  {Nagaosa}, \citenamefont {Sinova}, \citenamefont {Onoda}, \citenamefont
  {MacDonald},\ and\ \citenamefont {Ong}}]{RevModPhys.82.1539}%
  \BibitemOpen
  \bibfield  {author} {\bibinfo {author} {\bibfnamefont {N.}~\bibnamefont
  {Nagaosa}}, \bibinfo {author} {\bibfnamefont {J.}~\bibnamefont {Sinova}},
  \bibinfo {author} {\bibfnamefont {S.}~\bibnamefont {Onoda}}, \bibinfo
  {author} {\bibfnamefont {A.~H.}\ \bibnamefont {MacDonald}},\ and\ \bibinfo
  {author} {\bibfnamefont {N.~P.}\ \bibnamefont {Ong}},\ }\bibfield  {title}
  {\bibinfo {title} {Anomalous hall effect},\ }\href
  {https://doi.org/10.1103/RevModPhys.82.1539} {\bibfield  {journal} {\bibinfo
  {journal} {Rev. Mod. Phys.}\ }\textbf {\bibinfo {volume} {82}},\ \bibinfo
  {pages} {1539} (\bibinfo {year} {2010})}\BibitemShut {NoStop}%
\bibitem [{\citenamefont {Fang}\ \emph {et~al.}(2003)\citenamefont {Fang},
  \citenamefont {Nagaosa}, \citenamefont {Takahashi}, \citenamefont {Asamitsu},
  \citenamefont {Mathieu}, \citenamefont {Ogasawara}, \citenamefont {Yamada},
  \citenamefont {Kawasaki}, \citenamefont {Tokura},\ and\ \citenamefont
  {Terakura}}]{FangAHEScience}%
  \BibitemOpen
  \bibfield  {author} {\bibinfo {author} {\bibfnamefont {Z.}~\bibnamefont
  {Fang}}, \bibinfo {author} {\bibfnamefont {N.}~\bibnamefont {Nagaosa}},
  \bibinfo {author} {\bibfnamefont {K.~S.}\ \bibnamefont {Takahashi}}, \bibinfo
  {author} {\bibfnamefont {A.}~\bibnamefont {Asamitsu}}, \bibinfo {author}
  {\bibfnamefont {R.}~\bibnamefont {Mathieu}}, \bibinfo {author} {\bibfnamefont
  {T.}~\bibnamefont {Ogasawara}}, \bibinfo {author} {\bibfnamefont
  {H.}~\bibnamefont {Yamada}}, \bibinfo {author} {\bibfnamefont
  {M.}~\bibnamefont {Kawasaki}}, \bibinfo {author} {\bibfnamefont
  {Y.}~\bibnamefont {Tokura}},\ and\ \bibinfo {author} {\bibfnamefont
  {K.}~\bibnamefont {Terakura}},\ }\bibfield  {title} {\bibinfo {title} {The
  anomalous hall effect and magnetic monopoles in momentum space},\ }\href
  {https://doi.org/10.1126/science.1089408} {\bibfield  {journal} {\bibinfo
  {journal} {Science}\ }\textbf {\bibinfo {volume} {302}},\ \bibinfo {pages}
  {92} (\bibinfo {year} {2003})}\BibitemShut {NoStop}%
\bibitem [{\citenamefont {Jungwirth}\ \emph {et~al.}(2002)\citenamefont
  {Jungwirth}, \citenamefont {Niu},\ and\ \citenamefont
  {MacDonald}}]{AHEinFerroma}%
  \BibitemOpen
  \bibfield  {author} {\bibinfo {author} {\bibfnamefont {T.}~\bibnamefont
  {Jungwirth}}, \bibinfo {author} {\bibfnamefont {Q.}~\bibnamefont {Niu}},\
  and\ \bibinfo {author} {\bibfnamefont {A.~H.}\ \bibnamefont {MacDonald}},\
  }\bibfield  {title} {\bibinfo {title} {Anomalous hall effect in ferromagnetic
  semiconductors},\ }\href {https://doi.org/10.1103/PhysRevLett.88.207208}
  {\bibfield  {journal} {\bibinfo  {journal} {Phys. Rev. Lett.}\ }\textbf
  {\bibinfo {volume} {88}},\ \bibinfo {pages} {207208} (\bibinfo {year}
  {2002})}\BibitemShut {NoStop}%
\bibitem [{\citenamefont {Wan}\ \emph {et~al.}(2011)\citenamefont {Wan},
  \citenamefont {Turner}, \citenamefont {Vishwanath},\ and\ \citenamefont
  {Savrasov}}]{WanPhysRevB.83.205101}%
  \BibitemOpen
  \bibfield  {author} {\bibinfo {author} {\bibfnamefont {X.}~\bibnamefont
  {Wan}}, \bibinfo {author} {\bibfnamefont {A.~M.}\ \bibnamefont {Turner}},
  \bibinfo {author} {\bibfnamefont {A.}~\bibnamefont {Vishwanath}},\ and\
  \bibinfo {author} {\bibfnamefont {S.~Y.}\ \bibnamefont {Savrasov}},\
  }\bibfield  {title} {\bibinfo {title} {Topological semimetal and fermi-arc
  surface states in the electronic structure of pyrochlore iridates},\ }\href
  {https://doi.org/10.1103/PhysRevB.83.205101} {\bibfield  {journal} {\bibinfo
  {journal} {Phys. Rev. B}\ }\textbf {\bibinfo {volume} {83}},\ \bibinfo
  {pages} {205101} (\bibinfo {year} {2011})}\BibitemShut {NoStop}%
\bibitem [{\citenamefont {Soluyanov}\ \emph {et~al.}(2015)\citenamefont
  {Soluyanov}, \citenamefont {Gresch}, \citenamefont {Wang}, \citenamefont
  {Wu}, \citenamefont {Troyer}, \citenamefont {Dai},\ and\ \citenamefont
  {Bernevig}}]{Type2Weyl}%
  \BibitemOpen
  \bibfield  {author} {\bibinfo {author} {\bibfnamefont {A.~A.}\ \bibnamefont
  {Soluyanov}}, \bibinfo {author} {\bibfnamefont {D.}~\bibnamefont {Gresch}},
  \bibinfo {author} {\bibfnamefont {Z.}~\bibnamefont {Wang}}, \bibinfo {author}
  {\bibfnamefont {Q.}~\bibnamefont {Wu}}, \bibinfo {author} {\bibfnamefont
  {M.}~\bibnamefont {Troyer}}, \bibinfo {author} {\bibfnamefont
  {X.}~\bibnamefont {Dai}},\ and\ \bibinfo {author} {\bibfnamefont {B.~A.}\
  \bibnamefont {Bernevig}},\ }\bibfield  {title} {\bibinfo {title} {Type-ii
  weyl semimetals},\ }\href {https://doi.org/10.1038/nature15768} {\bibfield
  {journal} {\bibinfo  {journal} {Nature}\ }\textbf {\bibinfo {volume} {527}},\
  \bibinfo {pages} {495 EP } (\bibinfo {year} {2015})}\BibitemShut {NoStop}%
\bibitem [{\citenamefont {Wang}\ \emph {et~al.}(2012)\citenamefont {Wang},
  \citenamefont {Sun}, \citenamefont {Chen}, \citenamefont {Franchini},
  \citenamefont {Xu}, \citenamefont {Weng}, \citenamefont {Dai},\ and\
  \citenamefont {Fang}}]{WangNa3Bi}%
  \BibitemOpen
  \bibfield  {author} {\bibinfo {author} {\bibfnamefont {Z.}~\bibnamefont
  {Wang}}, \bibinfo {author} {\bibfnamefont {Y.}~\bibnamefont {Sun}}, \bibinfo
  {author} {\bibfnamefont {X.-Q.}\ \bibnamefont {Chen}}, \bibinfo {author}
  {\bibfnamefont {C.}~\bibnamefont {Franchini}}, \bibinfo {author}
  {\bibfnamefont {G.}~\bibnamefont {Xu}}, \bibinfo {author} {\bibfnamefont
  {H.}~\bibnamefont {Weng}}, \bibinfo {author} {\bibfnamefont {X.}~\bibnamefont
  {Dai}},\ and\ \bibinfo {author} {\bibfnamefont {Z.}~\bibnamefont {Fang}},\
  }\bibfield  {title} {\bibinfo {title} {Dirac semimetal and topological phase
  transitions in ${A}_{3}$bi ($a=\text{Na}$, k, rb)},\ }\href
  {https://doi.org/10.1103/PhysRevB.85.195320} {\bibfield  {journal} {\bibinfo
  {journal} {Phys. Rev. B}\ }\textbf {\bibinfo {volume} {85}},\ \bibinfo
  {pages} {195320} (\bibinfo {year} {2012})}\BibitemShut {NoStop}%
\bibitem [{\citenamefont {Liu}\ \emph {et~al.}(2014)\citenamefont {Liu},
  \citenamefont {Jiang}, \citenamefont {Zhou}, \citenamefont {Wang},
  \citenamefont {Zhang}, \citenamefont {Weng}, \citenamefont {Prabhakaran},
  \citenamefont {Mo}, \citenamefont {Peng}, \citenamefont {Dudin},
  \citenamefont {Kim}, \citenamefont {Hoesch}, \citenamefont {Fang},
  \citenamefont {Dai}, \citenamefont {Shen}, \citenamefont {Feng},
  \citenamefont {Hussain},\ and\ \citenamefont {Chen}}]{Liu2014Dirac}%
  \BibitemOpen
  \bibfield  {author} {\bibinfo {author} {\bibfnamefont {Z.~K.}\ \bibnamefont
  {Liu}}, \bibinfo {author} {\bibfnamefont {J.}~\bibnamefont {Jiang}}, \bibinfo
  {author} {\bibfnamefont {B.}~\bibnamefont {Zhou}}, \bibinfo {author}
  {\bibfnamefont {Z.~J.}\ \bibnamefont {Wang}}, \bibinfo {author}
  {\bibfnamefont {Y.}~\bibnamefont {Zhang}}, \bibinfo {author} {\bibfnamefont
  {H.~M.}\ \bibnamefont {Weng}}, \bibinfo {author} {\bibfnamefont
  {D.}~\bibnamefont {Prabhakaran}}, \bibinfo {author} {\bibfnamefont {S.-K.}\
  \bibnamefont {Mo}}, \bibinfo {author} {\bibfnamefont {H.}~\bibnamefont
  {Peng}}, \bibinfo {author} {\bibfnamefont {P.}~\bibnamefont {Dudin}},
  \bibinfo {author} {\bibfnamefont {T.}~\bibnamefont {Kim}}, \bibinfo {author}
  {\bibfnamefont {M.}~\bibnamefont {Hoesch}}, \bibinfo {author} {\bibfnamefont
  {Z.}~\bibnamefont {Fang}}, \bibinfo {author} {\bibfnamefont {X.}~\bibnamefont
  {Dai}}, \bibinfo {author} {\bibfnamefont {Z.~X.}\ \bibnamefont {Shen}},
  \bibinfo {author} {\bibfnamefont {D.~L.}\ \bibnamefont {Feng}}, \bibinfo
  {author} {\bibfnamefont {Z.}~\bibnamefont {Hussain}},\ and\ \bibinfo {author}
  {\bibfnamefont {Y.~L.}\ \bibnamefont {Chen}},\ }\bibfield  {title} {\bibinfo
  {title} {A stable three-dimensional topological dirac semimetal cd3as2},\
  }\href {https://doi.org/10.1038/nmat3990} {\bibfield  {journal} {\bibinfo
  {journal} {Nature Materials}\ }\textbf {\bibinfo {volume} {13}},\ \bibinfo
  {pages} {677 EP } (\bibinfo {year} {2014})}\BibitemShut {NoStop}%
\bibitem [{\citenamefont {Weng}\ \emph {et~al.}(2015)\citenamefont {Weng},
  \citenamefont {Fang}, \citenamefont {Fang}, \citenamefont {Bernevig},\ and\
  \citenamefont {Dai}}]{WengTaAsPRX}%
  \BibitemOpen
  \bibfield  {author} {\bibinfo {author} {\bibfnamefont {H.}~\bibnamefont
  {Weng}}, \bibinfo {author} {\bibfnamefont {C.}~\bibnamefont {Fang}}, \bibinfo
  {author} {\bibfnamefont {Z.}~\bibnamefont {Fang}}, \bibinfo {author}
  {\bibfnamefont {B.~A.}\ \bibnamefont {Bernevig}},\ and\ \bibinfo {author}
  {\bibfnamefont {X.}~\bibnamefont {Dai}},\ }\bibfield  {title} {\bibinfo
  {title} {Weyl semimetal phase in noncentrosymmetric transition-metal
  monophosphides},\ }\href {https://doi.org/10.1103/PhysRevX.5.011029}
  {\bibfield  {journal} {\bibinfo  {journal} {Phys. Rev. X}\ }\textbf {\bibinfo
  {volume} {5}},\ \bibinfo {pages} {011029} (\bibinfo {year}
  {2015})}\BibitemShut {NoStop}%
\bibitem [{\citenamefont {Lv}\ \emph {et~al.}(2015)\citenamefont {Lv},
  \citenamefont {Weng}, \citenamefont {Fu}, \citenamefont {Wang}, \citenamefont
  {Miao}, \citenamefont {Ma}, \citenamefont {Richard}, \citenamefont {Huang},
  \citenamefont {Zhao}, \citenamefont {Chen}, \citenamefont {Fang},
  \citenamefont {Dai}, \citenamefont {Qian},\ and\ \citenamefont
  {Ding}}]{ExpTaAsPRX}%
  \BibitemOpen
  \bibfield  {author} {\bibinfo {author} {\bibfnamefont {B.~Q.}\ \bibnamefont
  {Lv}}, \bibinfo {author} {\bibfnamefont {H.~M.}\ \bibnamefont {Weng}},
  \bibinfo {author} {\bibfnamefont {B.~B.}\ \bibnamefont {Fu}}, \bibinfo
  {author} {\bibfnamefont {X.~P.}\ \bibnamefont {Wang}}, \bibinfo {author}
  {\bibfnamefont {H.}~\bibnamefont {Miao}}, \bibinfo {author} {\bibfnamefont
  {J.}~\bibnamefont {Ma}}, \bibinfo {author} {\bibfnamefont {P.}~\bibnamefont
  {Richard}}, \bibinfo {author} {\bibfnamefont {X.~C.}\ \bibnamefont {Huang}},
  \bibinfo {author} {\bibfnamefont {L.~X.}\ \bibnamefont {Zhao}}, \bibinfo
  {author} {\bibfnamefont {G.~F.}\ \bibnamefont {Chen}}, \bibinfo {author}
  {\bibfnamefont {Z.}~\bibnamefont {Fang}}, \bibinfo {author} {\bibfnamefont
  {X.}~\bibnamefont {Dai}}, \bibinfo {author} {\bibfnamefont {T.}~\bibnamefont
  {Qian}},\ and\ \bibinfo {author} {\bibfnamefont {H.}~\bibnamefont {Ding}},\
  }\bibfield  {title} {\bibinfo {title} {Experimental discovery of weyl
  semimetal taas},\ }\href {https://doi.org/10.1103/PhysRevX.5.031013}
  {\bibfield  {journal} {\bibinfo  {journal} {Phys. Rev. X}\ }\textbf {\bibinfo
  {volume} {5}},\ \bibinfo {pages} {031013} (\bibinfo {year}
  {2015})}\BibitemShut {NoStop}%
\bibitem [{\citenamefont {Xu}\ \emph {et~al.}(2015)\citenamefont {Xu},
  \citenamefont {Belopolski}, \citenamefont {Alidoust}, \citenamefont
  {Neupane}, \citenamefont {Bian}, \citenamefont {Zhang}, \citenamefont
  {Sankar}, \citenamefont {Chang}, \citenamefont {Yuan}, \citenamefont {Lee},
  \citenamefont {Huang}, \citenamefont {Zheng}, \citenamefont {Ma},
  \citenamefont {Sanchez}, \citenamefont {Wang}, \citenamefont {Bansil},
  \citenamefont {Chou}, \citenamefont {Shibayev}, \citenamefont {Lin},
  \citenamefont {Jia},\ and\ \citenamefont {Hasan}}]{Xu613}%
  \BibitemOpen
  \bibfield  {author} {\bibinfo {author} {\bibfnamefont {S.-Y.}\ \bibnamefont
  {Xu}}, \bibinfo {author} {\bibfnamefont {I.}~\bibnamefont {Belopolski}},
  \bibinfo {author} {\bibfnamefont {N.}~\bibnamefont {Alidoust}}, \bibinfo
  {author} {\bibfnamefont {M.}~\bibnamefont {Neupane}}, \bibinfo {author}
  {\bibfnamefont {G.}~\bibnamefont {Bian}}, \bibinfo {author} {\bibfnamefont
  {C.}~\bibnamefont {Zhang}}, \bibinfo {author} {\bibfnamefont
  {R.}~\bibnamefont {Sankar}}, \bibinfo {author} {\bibfnamefont
  {G.}~\bibnamefont {Chang}}, \bibinfo {author} {\bibfnamefont
  {Z.}~\bibnamefont {Yuan}}, \bibinfo {author} {\bibfnamefont {C.-C.}\
  \bibnamefont {Lee}}, \bibinfo {author} {\bibfnamefont {S.-M.}\ \bibnamefont
  {Huang}}, \bibinfo {author} {\bibfnamefont {H.}~\bibnamefont {Zheng}},
  \bibinfo {author} {\bibfnamefont {J.}~\bibnamefont {Ma}}, \bibinfo {author}
  {\bibfnamefont {D.~S.}\ \bibnamefont {Sanchez}}, \bibinfo {author}
  {\bibfnamefont {B.}~\bibnamefont {Wang}}, \bibinfo {author} {\bibfnamefont
  {A.}~\bibnamefont {Bansil}}, \bibinfo {author} {\bibfnamefont
  {F.}~\bibnamefont {Chou}}, \bibinfo {author} {\bibfnamefont {P.~P.}\
  \bibnamefont {Shibayev}}, \bibinfo {author} {\bibfnamefont {H.}~\bibnamefont
  {Lin}}, \bibinfo {author} {\bibfnamefont {S.}~\bibnamefont {Jia}},\ and\
  \bibinfo {author} {\bibfnamefont {M.~Z.}\ \bibnamefont {Hasan}},\ }\bibfield
  {title} {\bibinfo {title} {Discovery of a weyl fermion semimetal and
  topological fermi arcs},\ }\href {https://doi.org/10.1126/science.aaa9297}
  {\bibfield  {journal} {\bibinfo  {journal} {Science}\ }\textbf {\bibinfo
  {volume} {349}},\ \bibinfo {pages} {613} (\bibinfo {year}
  {2015})}\BibitemShut {NoStop}%
\bibitem [{\citenamefont {Kim}\ \emph {et~al.}(2015)\citenamefont {Kim},
  \citenamefont {Wieder}, \citenamefont {Kane},\ and\ \citenamefont
  {Rappe}}]{KimDiracLineNodes}%
  \BibitemOpen
  \bibfield  {author} {\bibinfo {author} {\bibfnamefont {Y.}~\bibnamefont
  {Kim}}, \bibinfo {author} {\bibfnamefont {B.~J.}\ \bibnamefont {Wieder}},
  \bibinfo {author} {\bibfnamefont {C.~L.}\ \bibnamefont {Kane}},\ and\
  \bibinfo {author} {\bibfnamefont {A.~M.}\ \bibnamefont {Rappe}},\ }\bibfield
  {title} {\bibinfo {title} {Dirac line nodes in inversion-symmetric
  crystals},\ }\href {https://doi.org/10.1103/PhysRevLett.115.036806}
  {\bibfield  {journal} {\bibinfo  {journal} {Phys. Rev. Lett.}\ }\textbf
  {\bibinfo {volume} {115}},\ \bibinfo {pages} {036806} (\bibinfo {year}
  {2015})}\BibitemShut {NoStop}%
\bibitem [{\citenamefont {Schoop}\ \emph {et~al.}(2016)\citenamefont {Schoop},
  \citenamefont {Ali}, \citenamefont {Stra{\ss}er}, \citenamefont {Topp},
  \citenamefont {Varykhalov}, \citenamefont {Marchenko}, \citenamefont
  {Duppel}, \citenamefont {Parkin}, \citenamefont {Lotsch},\ and\ \citenamefont
  {Ast}}]{Schoop2016DiracLineNode}%
  \BibitemOpen
  \bibfield  {author} {\bibinfo {author} {\bibfnamefont {L.~M.}\ \bibnamefont
  {Schoop}}, \bibinfo {author} {\bibfnamefont {M.~N.}\ \bibnamefont {Ali}},
  \bibinfo {author} {\bibfnamefont {C.}~\bibnamefont {Stra{\ss}er}}, \bibinfo
  {author} {\bibfnamefont {A.}~\bibnamefont {Topp}}, \bibinfo {author}
  {\bibfnamefont {A.}~\bibnamefont {Varykhalov}}, \bibinfo {author}
  {\bibfnamefont {D.}~\bibnamefont {Marchenko}}, \bibinfo {author}
  {\bibfnamefont {V.}~\bibnamefont {Duppel}}, \bibinfo {author} {\bibfnamefont
  {S.~S.~P.}\ \bibnamefont {Parkin}}, \bibinfo {author} {\bibfnamefont {B.~V.}\
  \bibnamefont {Lotsch}},\ and\ \bibinfo {author} {\bibfnamefont {C.~R.}\
  \bibnamefont {Ast}},\ }\bibfield  {title} {\bibinfo {title} {Dirac cone
  protected by non-symmorphic symmetry and three-dimensional dirac line node in
  zrsis},\ }\href {https://doi.org/10.1038/ncomms11696} {\bibfield  {journal}
  {\bibinfo  {journal} {Nature Communications}\ }\textbf {\bibinfo {volume}
  {7}},\ \bibinfo {pages} {11696} (\bibinfo {year} {2016})}\BibitemShut
  {NoStop}%
\bibitem [{\citenamefont {Armitage}\ \emph {et~al.}(2018)\citenamefont
  {Armitage}, \citenamefont {Mele},\ and\ \citenamefont
  {Vishwanath}}]{RevModPhys.90.015001}%
  \BibitemOpen
  \bibfield  {author} {\bibinfo {author} {\bibfnamefont {N.~P.}\ \bibnamefont
  {Armitage}}, \bibinfo {author} {\bibfnamefont {E.~J.}\ \bibnamefont {Mele}},\
  and\ \bibinfo {author} {\bibfnamefont {A.}~\bibnamefont {Vishwanath}},\
  }\bibfield  {title} {\bibinfo {title} {Weyl and dirac semimetals in
  three-dimensional solids},\ }\href
  {https://doi.org/10.1103/RevModPhys.90.015001} {\bibfield  {journal}
  {\bibinfo  {journal} {Rev. Mod. Phys.}\ }\textbf {\bibinfo {volume} {90}},\
  \bibinfo {pages} {015001} (\bibinfo {year} {2018})}\BibitemShut {NoStop}%
\bibitem [{\citenamefont {Fukushima}\ \emph {et~al.}(2008)\citenamefont
  {Fukushima}, \citenamefont {Kharzeev},\ and\ \citenamefont
  {Warringa}}]{fukushima2008chiral}%
  \BibitemOpen
  \bibfield  {author} {\bibinfo {author} {\bibfnamefont {K.}~\bibnamefont
  {Fukushima}}, \bibinfo {author} {\bibfnamefont {D.~E.}\ \bibnamefont
  {Kharzeev}},\ and\ \bibinfo {author} {\bibfnamefont {H.~J.}\ \bibnamefont
  {Warringa}},\ }\bibfield  {title} {\bibinfo {title} {Chiral magnetic
  effect},\ }\href {https://doi.org/10.1103/PhysRevD.78.074033} {\bibfield
  {journal} {\bibinfo  {journal} {Phys. Rev. D}\ }\textbf {\bibinfo {volume}
  {78}},\ \bibinfo {pages} {074033} (\bibinfo {year} {2008})}\BibitemShut
  {NoStop}%
\bibitem [{\citenamefont {Li}\ \emph {et~al.}(2016)\citenamefont {Li},
  \citenamefont {Kharzeev}, \citenamefont {Zhang}, \citenamefont {Huang},
  \citenamefont {Pletikosic}, \citenamefont {Fedorov}, \citenamefont {Zhong},
  \citenamefont {Schneeloch}, \citenamefont {Gu},\ and\ \citenamefont
  {Valla}}]{CMEZrTr5}%
  \BibitemOpen
  \bibfield  {author} {\bibinfo {author} {\bibfnamefont {Q.}~\bibnamefont
  {Li}}, \bibinfo {author} {\bibfnamefont {D.~E.}\ \bibnamefont {Kharzeev}},
  \bibinfo {author} {\bibfnamefont {C.}~\bibnamefont {Zhang}}, \bibinfo
  {author} {\bibfnamefont {Y.}~\bibnamefont {Huang}}, \bibinfo {author}
  {\bibfnamefont {I.}~\bibnamefont {Pletikosic}}, \bibinfo {author}
  {\bibfnamefont {A.~.~V.}\ \bibnamefont {Fedorov}}, \bibinfo {author}
  {\bibfnamefont {R.~.~D.}\ \bibnamefont {Zhong}}, \bibinfo {author}
  {\bibfnamefont {J.~.~A.}\ \bibnamefont {Schneeloch}}, \bibinfo {author}
  {\bibfnamefont {G.~.~D.}\ \bibnamefont {Gu}},\ and\ \bibinfo {author}
  {\bibfnamefont {T.}~\bibnamefont {Valla}},\ }\bibfield  {title} {\bibinfo
  {title} {Chiral magnetic effect in zrte5},\ }\href
  {https://doi.org/10.1038/nphys3648} {\bibfield  {journal} {\bibinfo
  {journal} {Nature Physics}\ }\textbf {\bibinfo {volume} {12}},\ \bibinfo
  {pages} {550 EP } (\bibinfo {year} {2016})}\BibitemShut {NoStop}%
\bibitem [{\citenamefont {Son}\ and\ \citenamefont {Yamamoto}(2012)}]{CMEson}%
  \BibitemOpen
  \bibfield  {author} {\bibinfo {author} {\bibfnamefont {D.~T.}\ \bibnamefont
  {Son}}\ and\ \bibinfo {author} {\bibfnamefont {N.}~\bibnamefont {Yamamoto}},\
  }\bibfield  {title} {\bibinfo {title} {Berry curvature, triangle anomalies,
  and the chiral magnetic effect in fermi liquids},\ }\href
  {https://doi.org/10.1103/PhysRevLett.109.181602} {\bibfield  {journal}
  {\bibinfo  {journal} {Phys. Rev. Lett.}\ }\textbf {\bibinfo {volume} {109}},\
  \bibinfo {pages} {181602} (\bibinfo {year} {2012})}\BibitemShut {NoStop}%
\bibitem [{\citenamefont {Nielsen}\ and\ \citenamefont
  {Ninomiya}(1983)}]{abj_1983}%
  \BibitemOpen
  \bibfield  {author} {\bibinfo {author} {\bibfnamefont {H.~B.}\ \bibnamefont
  {Nielsen}}\ and\ \bibinfo {author} {\bibfnamefont {M.}~\bibnamefont
  {Ninomiya}},\ }\bibfield  {title} {\bibinfo {title} {The
  {Adler}-{Bell}-{Jackiw} anomaly and {Weyl} fermions in a crystal},\ }\href
  {https://doi.org/10.1016/0370-2693(83)91529-0} {\bibfield  {journal}
  {\bibinfo  {journal} {Physics Letters B}\ }\textbf {\bibinfo {volume}
  {130}},\ \bibinfo {pages} {389} (\bibinfo {year} {1983})}\BibitemShut
  {NoStop}%
\bibitem [{\citenamefont {Huang}\ \emph {et~al.}(2015)\citenamefont {Huang},
  \citenamefont {Zhao}, \citenamefont {Long}, \citenamefont {Wang},
  \citenamefont {Chen}, \citenamefont {Yang}, \citenamefont {Liang},
  \citenamefont {Xue}, \citenamefont {Weng}, \citenamefont {Fang},
  \citenamefont {Dai},\ and\ \citenamefont {Chen}}]{PhysRevX.5.031023}%
  \BibitemOpen
  \bibfield  {author} {\bibinfo {author} {\bibfnamefont {X.}~\bibnamefont
  {Huang}}, \bibinfo {author} {\bibfnamefont {L.}~\bibnamefont {Zhao}},
  \bibinfo {author} {\bibfnamefont {Y.}~\bibnamefont {Long}}, \bibinfo {author}
  {\bibfnamefont {P.}~\bibnamefont {Wang}}, \bibinfo {author} {\bibfnamefont
  {D.}~\bibnamefont {Chen}}, \bibinfo {author} {\bibfnamefont {Z.}~\bibnamefont
  {Yang}}, \bibinfo {author} {\bibfnamefont {H.}~\bibnamefont {Liang}},
  \bibinfo {author} {\bibfnamefont {M.}~\bibnamefont {Xue}}, \bibinfo {author}
  {\bibfnamefont {H.}~\bibnamefont {Weng}}, \bibinfo {author} {\bibfnamefont
  {Z.}~\bibnamefont {Fang}}, \bibinfo {author} {\bibfnamefont {X.}~\bibnamefont
  {Dai}},\ and\ \bibinfo {author} {\bibfnamefont {G.}~\bibnamefont {Chen}},\
  }\bibfield  {title} {\bibinfo {title} {Observation of the
  chiral-anomaly-induced negative magnetoresistance in 3d weyl semimetal
  taas},\ }\href {https://doi.org/10.1103/PhysRevX.5.031023} {\bibfield
  {journal} {\bibinfo  {journal} {Phys. Rev. X}\ }\textbf {\bibinfo {volume}
  {5}},\ \bibinfo {pages} {031023} (\bibinfo {year} {2015})}\BibitemShut
  {NoStop}%
\bibitem [{\citenamefont {Son}\ and\ \citenamefont
  {Spivak}(2013)}]{son_NMR_2013}%
  \BibitemOpen
  \bibfield  {author} {\bibinfo {author} {\bibfnamefont {D.~T.}\ \bibnamefont
  {Son}}\ and\ \bibinfo {author} {\bibfnamefont {B.~Z.}\ \bibnamefont
  {Spivak}},\ }\bibfield  {title} {\bibinfo {title} {Chiral anomaly and
  classical negative magnetoresistance of {Weyl} metals},\ }\href
  {https://doi.org/10.1103/PhysRevB.88.104412} {\bibfield  {journal} {\bibinfo
  {journal} {Physical Review B}\ }\textbf {\bibinfo {volume} {88}},\ \bibinfo
  {pages} {104412} (\bibinfo {year} {2013})}\BibitemShut {NoStop}%
\bibitem [{\citenamefont {Xiao}\ \emph {et~al.}(2010)\citenamefont {Xiao},
  \citenamefont {Chang},\ and\ \citenamefont {Niu}}]{RevModPhys.Niu}%
  \BibitemOpen
  \bibfield  {author} {\bibinfo {author} {\bibfnamefont {D.}~\bibnamefont
  {Xiao}}, \bibinfo {author} {\bibfnamefont {M.-C.}\ \bibnamefont {Chang}},\
  and\ \bibinfo {author} {\bibfnamefont {Q.}~\bibnamefont {Niu}},\ }\bibfield
  {title} {\bibinfo {title} {Berry phase effects on electronic properties},\
  }\href {https://doi.org/10.1103/RevModPhys.82.1959} {\bibfield  {journal}
  {\bibinfo  {journal} {Rev. Mod. Phys.}\ }\textbf {\bibinfo {volume} {82}},\
  \bibinfo {pages} {1959} (\bibinfo {year} {2010})}\BibitemShut {NoStop}%
\bibitem [{\citenamefont {Winkler}()}]{winkler_spin--orbit_2003}%
  \BibitemOpen
  \bibfield  {author} {\bibinfo {author} {\bibfnamefont {R.}~\bibnamefont
  {Winkler}},\ }\href {https://doi.org/10.1007/b13586} {\emph {\bibinfo {title}
  {Spin-Orbit Coupling Effects in Two-Dimensional Electron and Hole
  Systems}}},\ \bibinfo {series} {Springer Tracts in Modern Physics}, Vol.\
  \bibinfo {volume} {191}\ (\bibinfo  {publisher} {Springer Berlin
  Heidelberg})\ pp.\ \bibinfo {pages} {9--11,201--205}\BibitemShut {NoStop}%
\bibitem [{\citenamefont {L{\"o}wdin}(1951)}]{lowdin1951note}%
  \BibitemOpen
  \bibfield  {author} {\bibinfo {author} {\bibfnamefont {P.-O.}\ \bibnamefont
  {L{\"o}wdin}},\ }\bibfield  {title} {\bibinfo {title} {A note on the
  quantum-mechanical perturbation theory},\ }\href
  {http://dx.doi.org/10.1063/1.1748067} {\bibfield  {journal} {\bibinfo
  {journal} {The Journal of Chemical Physics}\ }\textbf {\bibinfo {volume}
  {19}},\ \bibinfo {pages} {1396} (\bibinfo {year} {1951})}\BibitemShut
  {NoStop}%
\bibitem [{\citenamefont {Weng}\ \emph {et~al.}(2014)\citenamefont {Weng},
  \citenamefont {Dai},\ and\ \citenamefont
  {Fang}}]{weng_transition-metal_2014}%
  \BibitemOpen
  \bibfield  {author} {\bibinfo {author} {\bibfnamefont {H.}~\bibnamefont
  {Weng}}, \bibinfo {author} {\bibfnamefont {X.}~\bibnamefont {Dai}},\ and\
  \bibinfo {author} {\bibfnamefont {Z.}~\bibnamefont {Fang}},\ }\bibfield
  {title} {\bibinfo {title} {Transition-metal pentatelluride
  $\mathrm{ZrTe}{}_{5}$ and $\mathrm{HfTe}{}_{5}$: A paradigm for large-gap
  quantum spin hall insulators},\ }\href
  {https://doi.org/10.1103/PhysRevX.4.011002} {\bibfield  {journal} {\bibinfo
  {journal} {Phys. Rev. X}\ }\textbf {\bibinfo {volume} {4}},\ \bibinfo {pages}
  {011002} (\bibinfo {year} {2014})}\BibitemShut {NoStop}%
\bibitem [{\citenamefont {Liu}\ \emph {et~al.}(2016)\citenamefont {Liu},
  \citenamefont {Yuan}, \citenamefont {Zhang}, \citenamefont {Jin},
  \citenamefont {Narayan}, \citenamefont {Luo}, \citenamefont {Chen},
  \citenamefont {Yang}, \citenamefont {Zou}, \citenamefont {Wu}, \citenamefont
  {Sanvito}, \citenamefont {Xia}, \citenamefont {Li}, \citenamefont {Wang},\
  and\ \citenamefont {Xiu}}]{Liu2016}%
  \BibitemOpen
  \bibfield  {author} {\bibinfo {author} {\bibfnamefont {Y.}~\bibnamefont
  {Liu}}, \bibinfo {author} {\bibfnamefont {X.}~\bibnamefont {Yuan}}, \bibinfo
  {author} {\bibfnamefont {C.}~\bibnamefont {Zhang}}, \bibinfo {author}
  {\bibfnamefont {Z.}~\bibnamefont {Jin}}, \bibinfo {author} {\bibfnamefont
  {A.}~\bibnamefont {Narayan}}, \bibinfo {author} {\bibfnamefont
  {C.}~\bibnamefont {Luo}}, \bibinfo {author} {\bibfnamefont {Z.}~\bibnamefont
  {Chen}}, \bibinfo {author} {\bibfnamefont {L.}~\bibnamefont {Yang}}, \bibinfo
  {author} {\bibfnamefont {J.}~\bibnamefont {Zou}}, \bibinfo {author}
  {\bibfnamefont {X.}~\bibnamefont {Wu}}, \bibinfo {author} {\bibfnamefont
  {S.}~\bibnamefont {Sanvito}}, \bibinfo {author} {\bibfnamefont
  {Z.}~\bibnamefont {Xia}}, \bibinfo {author} {\bibfnamefont {L.}~\bibnamefont
  {Li}}, \bibinfo {author} {\bibfnamefont {Z.}~\bibnamefont {Wang}},\ and\
  \bibinfo {author} {\bibfnamefont {F.}~\bibnamefont {Xiu}},\ }\bibfield
  {title} {\bibinfo {title} {Zeeman splitting and dynamical mass generation in
  dirac semimetal zrte5},\ }\href {https://doi.org/10.1038/ncomms12516}
  {\bibfield  {journal} {\bibinfo  {journal} {Nature Communications}\ }\textbf
  {\bibinfo {volume} {7}},\ \bibinfo {pages} {12516} (\bibinfo {year}
  {2016})}\BibitemShut {NoStop}%
\bibitem [{\citenamefont {Zheng}\ \emph {et~al.}(2016)\citenamefont {Zheng},
  \citenamefont {Lu}, \citenamefont {Zhu}, \citenamefont {Ning}, \citenamefont
  {Han}, \citenamefont {Zhang}, \citenamefont {Zhang}, \citenamefont {Xi},
  \citenamefont {Yang}, \citenamefont {Du}, \citenamefont {Yang}, \citenamefont
  {Zhang},\ and\ \citenamefont {Tian}}]{ZrTe5PhysRevB.93.115414}%
  \BibitemOpen
  \bibfield  {author} {\bibinfo {author} {\bibfnamefont {G.}~\bibnamefont
  {Zheng}}, \bibinfo {author} {\bibfnamefont {J.}~\bibnamefont {Lu}}, \bibinfo
  {author} {\bibfnamefont {X.}~\bibnamefont {Zhu}}, \bibinfo {author}
  {\bibfnamefont {W.}~\bibnamefont {Ning}}, \bibinfo {author} {\bibfnamefont
  {Y.}~\bibnamefont {Han}}, \bibinfo {author} {\bibfnamefont {H.}~\bibnamefont
  {Zhang}}, \bibinfo {author} {\bibfnamefont {J.}~\bibnamefont {Zhang}},
  \bibinfo {author} {\bibfnamefont {C.}~\bibnamefont {Xi}}, \bibinfo {author}
  {\bibfnamefont {J.}~\bibnamefont {Yang}}, \bibinfo {author} {\bibfnamefont
  {H.}~\bibnamefont {Du}}, \bibinfo {author} {\bibfnamefont {K.}~\bibnamefont
  {Yang}}, \bibinfo {author} {\bibfnamefont {Y.}~\bibnamefont {Zhang}},\ and\
  \bibinfo {author} {\bibfnamefont {M.}~\bibnamefont {Tian}},\ }\bibfield
  {title} {\bibinfo {title} {Transport evidence for the three-dimensional dirac
  semimetal phase in $\mathrm{ZrT}{\mathrm{e}}_{5}$},\ }\href
  {https://doi.org/10.1103/PhysRevB.93.115414} {\bibfield  {journal} {\bibinfo
  {journal} {Phys. Rev. B}\ }\textbf {\bibinfo {volume} {93}},\ \bibinfo
  {pages} {115414} (\bibinfo {year} {2016})}\BibitemShut {NoStop}%
\bibitem [{\citenamefont {Kamm}\ \emph {et~al.}(1985)\citenamefont {Kamm},
  \citenamefont {Gillespie}, \citenamefont {Ehrlich}, \citenamefont {Wieting},\
  and\ \citenamefont {Levy}}]{ZrTe5PhysRevB.31.7617}%
  \BibitemOpen
  \bibfield  {author} {\bibinfo {author} {\bibfnamefont {G.~N.}\ \bibnamefont
  {Kamm}}, \bibinfo {author} {\bibfnamefont {D.~J.}\ \bibnamefont {Gillespie}},
  \bibinfo {author} {\bibfnamefont {A.~C.}\ \bibnamefont {Ehrlich}}, \bibinfo
  {author} {\bibfnamefont {T.~J.}\ \bibnamefont {Wieting}},\ and\ \bibinfo
  {author} {\bibfnamefont {F.}~\bibnamefont {Levy}},\ }\bibfield  {title}
  {\bibinfo {title} {Fermi surface, effective masses, and dingle temperatures
  of ${\mathrm{zrte}}_{5}$ as derived from the shubnikov--de haas effect},\
  }\href {https://doi.org/10.1103/PhysRevB.31.7617} {\bibfield  {journal}
  {\bibinfo  {journal} {Phys. Rev. B}\ }\textbf {\bibinfo {volume} {31}},\
  \bibinfo {pages} {7617} (\bibinfo {year} {1985})}\BibitemShut {NoStop}%
\bibitem [{\citenamefont {Wang}\ \emph {et~al.}(2018)\citenamefont {Wang},
  \citenamefont {Niu}, \citenamefont {Yan}, \citenamefont {Li}, \citenamefont
  {Bi}, \citenamefont {Yao}, \citenamefont {Yu},\ and\ \citenamefont
  {Wu}}]{wang_vanishing_2018}%
  \BibitemOpen
  \bibfield  {author} {\bibinfo {author} {\bibfnamefont {J.}~\bibnamefont
  {Wang}}, \bibinfo {author} {\bibfnamefont {J.}~\bibnamefont {Niu}}, \bibinfo
  {author} {\bibfnamefont {B.}~\bibnamefont {Yan}}, \bibinfo {author}
  {\bibfnamefont {X.}~\bibnamefont {Li}}, \bibinfo {author} {\bibfnamefont
  {R.}~\bibnamefont {Bi}}, \bibinfo {author} {\bibfnamefont {Y.}~\bibnamefont
  {Yao}}, \bibinfo {author} {\bibfnamefont {D.}~\bibnamefont {Yu}},\ and\
  \bibinfo {author} {\bibfnamefont {X.}~\bibnamefont {Wu}},\ }\bibfield
  {title} {\bibinfo {title} {Vanishing quantum oscillations in dirac semimetal
  zrte5},\ }\href {https://doi.org/10.1073/pnas.1804958115} {\bibfield
  {journal} {\bibinfo  {journal} {Proceedings of the National Academy of
  Sciences}\ }\textbf {\bibinfo {volume} {115}},\ \bibinfo {pages} {9145}
  (\bibinfo {year} {2018})}\BibitemShut {NoStop}%
\bibitem [{\citenamefont {Zhao}\ \emph {et~al.}(2016)\citenamefont {Zhao},
  \citenamefont {Schnyder},\ and\ \citenamefont {Wang}}]{PTClassification}%
  \BibitemOpen
  \bibfield  {author} {\bibinfo {author} {\bibfnamefont {Y.~X.}\ \bibnamefont
  {Zhao}}, \bibinfo {author} {\bibfnamefont {A.~P.}\ \bibnamefont {Schnyder}},\
  and\ \bibinfo {author} {\bibfnamefont {Z.~D.}\ \bibnamefont {Wang}},\
  }\bibfield  {title} {\bibinfo {title} {Unified theory of $pt$ and $cp$
  invariant topological metals and nodal superconductors},\ }\href
  {https://doi.org/10.1103/PhysRevLett.116.156402} {\bibfield  {journal}
  {\bibinfo  {journal} {Phys. Rev. Lett.}\ }\textbf {\bibinfo {volume} {116}},\
  \bibinfo {pages} {156402} (\bibinfo {year} {2016})}\BibitemShut {NoStop}%
\bibitem [{\citenamefont {Burkov}(2014)}]{burkov_CA}%
  \BibitemOpen
  \bibfield  {author} {\bibinfo {author} {\bibfnamefont {A.~A.}\ \bibnamefont
  {Burkov}},\ }\bibfield  {title} {\bibinfo {title} {Chiral anomaly and
  diffusive magnetotransport in weyl metals},\ }\href
  {https://doi.org/10.1103/PhysRevLett.113.247203} {\bibfield  {journal}
  {\bibinfo  {journal} {Phys. Rev. Lett.}\ }\textbf {\bibinfo {volume} {113}},\
  \bibinfo {pages} {247203} (\bibinfo {year} {2014})}\BibitemShut {NoStop}%
\bibitem [{\citenamefont {Burkov}(2015)}]{burkov_NMR}%
  \BibitemOpen
  \bibfield  {author} {\bibinfo {author} {\bibfnamefont {A.~A.}\ \bibnamefont
  {Burkov}},\ }\bibfield  {title} {\bibinfo {title} {Negative longitudinal
  magnetoresistance in {Dirac} and {Weyl} metals},\ }\href
  {https://doi.org/10.1103/PhysRevB.91.245157} {\bibfield  {journal} {\bibinfo
  {journal} {Physical Review B}\ }\textbf {\bibinfo {volume} {91}},\ \bibinfo
  {pages} {245157} (\bibinfo {year} {2015})}\BibitemShut {NoStop}%
\bibitem [{\citenamefont {Xiong}\ \emph {et~al.}(2015)\citenamefont {Xiong},
  \citenamefont {Kushwaha}, \citenamefont {Liang}, \citenamefont {Krizan},
  \citenamefont {Hirschberger}, \citenamefont {Wang}, \citenamefont {Cava},\
  and\ \citenamefont {Ong}}]{Na3Bi_NMR}%
  \BibitemOpen
  \bibfield  {author} {\bibinfo {author} {\bibfnamefont {J.}~\bibnamefont
  {Xiong}}, \bibinfo {author} {\bibfnamefont {S.~K.}\ \bibnamefont {Kushwaha}},
  \bibinfo {author} {\bibfnamefont {T.}~\bibnamefont {Liang}}, \bibinfo
  {author} {\bibfnamefont {J.~W.}\ \bibnamefont {Krizan}}, \bibinfo {author}
  {\bibfnamefont {M.}~\bibnamefont {Hirschberger}}, \bibinfo {author}
  {\bibfnamefont {W.}~\bibnamefont {Wang}}, \bibinfo {author} {\bibfnamefont
  {R.~J.}\ \bibnamefont {Cava}},\ and\ \bibinfo {author} {\bibfnamefont
  {N.~P.}\ \bibnamefont {Ong}},\ }\bibfield  {title} {\bibinfo {title}
  {Evidence for the chiral anomaly in the {Dirac} semimetal {Na}3bi},\ }\href
  {https://doi.org/10.1126/science.aac6089} {\bibfield  {journal} {\bibinfo
  {journal} {Science}\ }\textbf {\bibinfo {volume} {350}},\ \bibinfo {pages}
  {413} (\bibinfo {year} {2015})}\BibitemShut {NoStop}%
\bibitem [{\citenamefont {Wang}\ \emph {et~al.}(2006)\citenamefont {Wang},
  \citenamefont {Yates}, \citenamefont {Souza},\ and\ \citenamefont
  {Vanderbilt}}]{PhysRevB.74.195118}%
  \BibitemOpen
  \bibfield  {author} {\bibinfo {author} {\bibfnamefont {X.}~\bibnamefont
  {Wang}}, \bibinfo {author} {\bibfnamefont {J.~R.}\ \bibnamefont {Yates}},
  \bibinfo {author} {\bibfnamefont {I.}~\bibnamefont {Souza}},\ and\ \bibinfo
  {author} {\bibfnamefont {D.}~\bibnamefont {Vanderbilt}},\ }\bibfield  {title}
  {\bibinfo {title} {Ab initio calculation of the anomalous hall conductivity
  by wannier interpolation},\ }\href
  {https://doi.org/10.1103/PhysRevB.74.195118} {\bibfield  {journal} {\bibinfo
  {journal} {Phys. Rev. B}\ }\textbf {\bibinfo {volume} {74}},\ \bibinfo
  {pages} {195118} (\bibinfo {year} {2006})}\BibitemShut {NoStop}%
\bibitem [{\citenamefont {Persson}(2015)}]{osti_1188980}%
  \BibitemOpen
  \bibfield  {author} {\bibinfo {author} {\bibfnamefont {K.}~\bibnamefont
  {Persson}},\ }\href {https://doi.org/10.17188/1188980} {\bibinfo {title}
  {Materials data on taas2 (sg:12) by materials project}} (\bibinfo {year}
  {2015}),\ \bibinfo {note} {an optional note}\BibitemShut {NoStop}%
\bibitem [{\citenamefont {Luo}\ \emph {et~al.}(2016)\citenamefont {Luo},
  \citenamefont {McDonald}, \citenamefont {Rosa}, \citenamefont {Scott},
  \citenamefont {Wakeham}, \citenamefont {Ghimire}, \citenamefont {Bauer},
  \citenamefont {Thompson},\ and\ \citenamefont
  {Ronning}}]{luo_anomalous_2016}%
  \BibitemOpen
  \bibfield  {author} {\bibinfo {author} {\bibfnamefont {Y.}~\bibnamefont
  {Luo}}, \bibinfo {author} {\bibfnamefont {R.~D.}\ \bibnamefont {McDonald}},
  \bibinfo {author} {\bibfnamefont {P.~F.~S.}\ \bibnamefont {Rosa}}, \bibinfo
  {author} {\bibfnamefont {B.}~\bibnamefont {Scott}}, \bibinfo {author}
  {\bibfnamefont {N.}~\bibnamefont {Wakeham}}, \bibinfo {author} {\bibfnamefont
  {N.~J.}\ \bibnamefont {Ghimire}}, \bibinfo {author} {\bibfnamefont {E.~D.}\
  \bibnamefont {Bauer}}, \bibinfo {author} {\bibfnamefont {J.~D.}\ \bibnamefont
  {Thompson}},\ and\ \bibinfo {author} {\bibfnamefont {F.}~\bibnamefont
  {Ronning}},\ }\bibfield  {title} {\bibinfo {title} {Anomalous electronic
  structure and magnetoresistance in taas2},\ }\href
  {https://doi.org/10.1038/srep27294} {\bibfield  {journal} {\bibinfo
  {journal} {Scientific Reports}\ }\textbf {\bibinfo {volume} {6}},\ \bibinfo
  {pages} {27294 EP } (\bibinfo {year} {2016})},\ \bibinfo {note}
  {article}\BibitemShut {NoStop}%
\bibitem [{\citenamefont {Yuan}\ \emph {et~al.}(2016)\citenamefont {Yuan},
  \citenamefont {Lu}, \citenamefont {Liu}, \citenamefont {Wang},\ and\
  \citenamefont {Jia}}]{yuan2016large}%
  \BibitemOpen
  \bibfield  {author} {\bibinfo {author} {\bibfnamefont {Z.}~\bibnamefont
  {Yuan}}, \bibinfo {author} {\bibfnamefont {H.}~\bibnamefont {Lu}}, \bibinfo
  {author} {\bibfnamefont {Y.}~\bibnamefont {Liu}}, \bibinfo {author}
  {\bibfnamefont {J.}~\bibnamefont {Wang}},\ and\ \bibinfo {author}
  {\bibfnamefont {S.}~\bibnamefont {Jia}},\ }\bibfield  {title} {\bibinfo
  {title} {Large magnetoresistance in compensated semimetals taas 2 and nbas
  2},\ }\href@noop {} {\bibfield  {journal} {\bibinfo  {journal} {Physical
  Review B}\ }\textbf {\bibinfo {volume} {93}},\ \bibinfo {pages} {184405}
  (\bibinfo {year} {2016})}\BibitemShut {NoStop}%
\bibitem [{\citenamefont {Sakurai}(1994)}]{sakurai}%
  \BibitemOpen
  \bibfield  {author} {\bibinfo {author} {\bibfnamefont {J.~J.}\ \bibnamefont
  {Sakurai}},\ }\href
  {http://gen.lib.rus.ec/book/index.php?md5=135C77B33B4D32D809C3E39335DD6AE2}
  {\emph {\bibinfo {title} {Modern quantum mechanics}}},\ \bibinfo {edition}
  {rev. ed}\ ed.\ (\bibinfo  {publisher} {Addison-Wesley Pub. Co},\ \bibinfo
  {year} {1994})\ p.\ \bibinfo {pages} {269}\BibitemShut {NoStop}%
\end{thebibliography}%

\end{document}